\documentclass[12pt]{article}

\AtBeginDocument{
\addtocontents{toc}{\normalsize}
}
\pdfoutput=1

\usepackage{amsmath,amssymb,amsfonts,amsthm,amscd,mathrsfs}
\usepackage{esvect}
\usepackage{xcolor}
\definecolor{darkblue}{rgb}{0.1,0.1,.7}
\usepackage{booktabs}
\usepackage[]{version}
\usepackage[]{graphicx}
\usepackage[]{latexsym}
\usepackage{geometry}
\usepackage[all,cmtip]{xy}

\usepackage[margin=10pt,font=small,labelfont=bf]{caption}
\usepackage{ifthen}
\usepackage{soul}
\usepackage{tikz}
\usetikzlibrary{shapes.geometric}
\tikzset{
  vtx/.style={
    circle,
    draw=blue,
    fill=blue,
    inner sep=1pt
  },
  wcirc/.style={
    circle,
    draw=white,
    fill=white,
    inner sep=2pt
  },
  bcirc/.style={
    circle,
    draw=black,
    fill=black,
    inner sep=1pt
  },
  dcirc/.style={
    circle,
    draw=blue,
    fill=blue,
    inner sep=1pt
  },
  rcirc/.style={
    circle,
    draw=red,
    fill=red,
    inner sep=1pt
  },
  phi/.style={
    thick
  },
  sigma/.style={
    thick,
    dashed
  },
  vl1/.style={
    thick,
    blue
  },
  vl2/.style={
    thick,
    dashed,
    blue
  },
  valign/.style={
    baseline={([yshift=-.55ex]current bounding box.center)}
  }
}
\usepackage{pgfplots}
\pgfplotsset{width=10cm,compat=1.9}

\usepackage{slashed}
\usepackage{array,mathrsfs,amsfonts,yfonts,dsfont,bbm,colonequals}
\usepackage[nodisplayskipstretch]{setspace}
\usepackage{dsfont}
\usepackage{cite}
\usepackage{xspace}
\usepackage{empheq}
\usepackage{extarrows}

\usepackage{xcolor}
\usepackage[colorlinks, linkcolor=darkblue, citecolor=darkblue, urlcolor=darkblue, linktocpage]{hyperref} 

\usepackage{subcaption}
\usepackage{empheq}
\usepackage[utf8]{inputenc}
\usepackage{setspace}
\usepackage{float, graphicx}

\numberwithin{equation}{section}

\newcommand{\lrni}{{\tt LR}n{\tt I}}
\newcommand{\DN}{\Delta_{\tt N}}
\newcommand{\DD}{\Delta_{\tt D}}

\newcommand{\cO}{\mathcal O}

\newcommand{\reef}[1]{(\ref{#1})}
\newcommand{\be}{\begin{equation}}
\newcommand{\ee}{\end{equation}}
\newcommand{\bea}{\begin{eqnarray}}
\newcommand{\eea}{\end{eqnarray}}
\newcommand{\ba}{\begin{equation}\begin{aligned}}
\newcommand{\ea}{\end{aligned}\end{equation}}

\newcommand{\boxeda}[1]{%
  \begin{empheq}[box=\fbox]{equation}
    \begin{aligned}
      #1
    \end{aligned}
  \end{empheq}
}

\newcommand*\pFqskip{8mu}
\catcode`,\active
\newcommand*\pFq{\begingroup
\catcode`\,\active
\def ,{\mskip\pFqskip\relax}%
\dopFq
}
\newcommand*\pFqt{\begingroup
\catcode`\,\active
\def ,{\mskip\pFqskip\relax}%
\dopFqt
}

\catcode`\,12
\def\dopFq#1#2#3#4#5{%
{}_{#1}F_{#2}\biggl[\genfrac..{0pt}{}{#3}{#4};#5\biggr]%
\endgroup
}
\catcode`\,12
\def\dopFqt#1#2#3#4#5{%
{}_{#1}\tilde F_{#2}\biggl[\genfrac..{0pt}{}{#3}{#4};#5\biggr]%
\endgroup
}

\newcommand{\eps}{\epsilon}

\newcommand{\ud}{\mathrm d}

\newcommand{\Df}{{\Delta_\phi}}

\def\a{\alpha}

\def\G{\Gamma}

\def\G{\Gamma}

\usepackage{cancel}

\begin{document}

\vspace*{-.6in} \thispagestyle{empty}

\vspace{1cm} {\Large
\begin{center}
    {\bf Interfaces in Long Range CFT$_1$}\\
\end{center}}
\vspace{1cm}

\begin{center}
{\bf Bastien Girault, Miguel F.~Paulos}\\[1cm] 
{
    \small
    {Laboratoire de Physique,  \\ \'Ecole Normale Sup\'erieure, \\
Universit{\'e} PSL, CNRS, Sorbonne Universit{\'e}, Universit{\'e} Paris Cit{\'e}, \\
24 rue Lhomond, F-75005 Paris, France}

    \normalsize
}

\end{center}

\begin{center}
{  \texttt{bastien.girault@ens.fr\ \ \  miguel.paulos@ens.fr} 
}
\\
\end{center}

\vspace{8mm}

\begin{abstract}
\vspace{2mm}
We consider interfaces in the theory of a single free scalar in AdS$_2$ defined by adding relevant operators along portions of the boundary. We begin with Neumann/Dirichlet interfaces, computing the associated scaling dimension and $g$-factor ratio; we also explain how arbitrary interface correlators can be computed numerically efficiently. Then we move on to interfaces between generalized free field and long-range interacting theories such as long range Ising, determining many quantities in perturbation theory. Finally we setup a bootstrap problem for four-point interface correlator. We use this to find tight bounds on the $g$-factor ratio as a function of the interface scaling dimension.
\end{abstract}
\vspace{2in}

\newpage

{
\setlength{\parskip}{0.05in}
\tableofcontents
\renewcommand{\baselinestretch}{1.0}\normalsize
}

\setlength{\parskip}{0.1in}
\setlength{\abovedisplayskip}{15pt}
\setlength{\belowdisplayskip}{15pt}
\setlength{\belowdisplayshortskip}{15pt}
\setlength{\abovedisplayshortskip}{15pt}

\newpage

\section{Introduction}
Conformal field theories offer us a plethora of observables determined by both localized and extended probes. In the former category we have the local operators of the theory for which we have a relatively detailed understanding: the set of operators is determined by state-operator correspondence to match the spectrum of the theory on the cylinder, and their interactions are controlled by a convergent OPE, satisfying strong constraints from crossing symmetry and unitarity. For extended probes we ask about the set of defects of various codimensions allowed in the theory, typically preserving some subset of the conformal group on their worldvolume. Here our understanding is much more limited, although some constraints can be derived in special cases: in 2d CFT Virasoro symmetry can allow us to say something about multiple codimension-1 boundary conditions, and in general dimension a growing body of literature has been able to derive a multitude of constraints on defect CFT data by considering correlators in the presence of an extended defect~\cite{Liendo:2012hy,Gaiotto:2013nva,Gliozzi:2015qsa,Lemos:2017vnx,Liendo:2019jpu,Barrat:2022psm,Mazac:2018biw,Meineri:2025qwz}.

Conformal theories in one-dimension\footnote{Such theories are non-local, i.e. have no stress-tensor and are sometimes called Conformal Theories.} offer an intriguing playground where the worlds of extended and local probes collide. This is because in this case, trivially, the only allowed extended defects have codimension zero, which therefore terminate at point like insertions which in many ways behave as local operators. This offers an opportunity to learn about defect dynamics by using well-developed techniques applicable to local correlation functions, such as the conformal bootstrap \cite{Rattazzi:2008pe,Poland:2018epd}.

This idea was recently explored by the authors of~\cite{Lanzetta:2025xfw} in the context of conformally invariant line defects embedded inside a higher-dimensional CFT$_D$. In general the CFT$_D$ will support various such defects, defined say by some kind of boundary condition along a line. Any given defect has a spectrum of local operators, whose correlators define those of a one-dimensional CFT. The idea now is that somewhere along a particular line defect defining some CFT$_{1}^a$, we can introduce regions where the choice of boundary condition is different, so that they behave as some other CFT$_1^b$. Such regions can be thought of as insertions of codimension-0 extended objects on the original CFT$_1$. At the end points of these regions we have pointwise insertions which act as defect changing operators, $\mu^{a/ b}$, or interfaces between these CFTs.

Since the $\mu^{a/ b}$ behave as local operators, it is possible to set up a bootstrap problem for correlators involving them, as well as other ordinary local operator insertions in $\cO^{a}$ or $\cO^{b}$. In particular, if one of the boundary conditions , say $a$, is simply the trivial one, then $\cO^a$ are simply the original operators in the CFT$_D$ and we can exploit any knowledge we already have about their OPE. For instance in reference~\cite{Lanzetta:2025xfw} this CFT$_3$ was the 3d Ising CFT, for which a number of results are known about its CFT data~\cite{Simmons-Duffin:2016wlq,Chang:2024whx}. This led to tight bounds on the OPE data of defect changing operators and their dimensions.

The goal of this paper is to begin an exploration of this approach applied to a broader class of 1d CFTs. To motivate this generalization, consider a line embedded in a CFT$_D$. Doing a Weyl transformation we can map the geometry to a space AdS$_2\times S_{D-2}$, with the line defect CFT living at the boundary of AdS$_2$. The point now is that any quantum field theory in AdS$_2\times \mathcal M$ produces such a 1d CFT. One way to produce extended defects in the CFT$_1$ is then to make different choices of boundary conditions for the bulk QFT along different portions of the AdS$_2$ boundary, or more generally, by adding relevant defect operators along regions of the boundary and tuning to conformality. A simple familiar example which we will explore in detail is to take a free scalar $\Phi$ in AdS$_2$ with negative mass squared so that it supports both Neumann and Dirichlet boundary conditions~\cite{Klebanov:1999tb}. Starting from Neumann we can add $\phi^2$ along an extended region to generate a flow to the Dirichlet case. Since the theory is free many computations can be done exactly~\cite{Melby-Thompson:2017aip}.

More interesting examples arise when we add interactions. This can be the case for an interacting bulk theory, or for a free bulk in the presence of suitable relevant operators at the boundary. The 1d long range Ising CFT is an example of the former, which arises by starting from a generalized free field with dimension below $1/4$ and adding $\phi^4$ interaction. This produces an interface between the long range model on one side and a free field on the other. We can now use the same ideas of~\cite{Lanzetta:2025xfw} described above, but where now the theory trivial line defect is exactly and completely known. 

The outline of this paper is as follows:
\begin{itemize}
\item In section \ref{sec:gffint} we describe our basic setup involving a free scalar field in AdS$_2$ with a bulk interface, separating regions with distinct mass terms. We explain how the energy of the interface may be computed as well as the spectrum of its excitations.
\item In section \ref{sec:ndint} we review and extend previous results Neumann/Dirichlet interfaces, i.e. where the bulk mass term is continuous but there are two distinct choices of boundary condition on the field.
\item In section \ref{sec:multi-interface-correlators} we explain how multi-interface Neumann/Dirichlet correlators may be determined by path integral methods, leading to a simple formula amenable to numerical computation.
\item Section \ref{sec:lrint} studies interfaces between generalized free fields and multicritical long range fixed points, computing interface dimensions and $g$-factors in perturbation theory.
\item In section \ref{sec:bootstrap} we do a bootstrap analysis of 4-interface correlators, corresponding to two generalized free regions and two interacting long-range Ising regions. After a brief perturbative analytic analysis we move on to derive various bounds on $g$-factors and interface scaling dimension numerically.
\item We conclude in section \ref{sec:conclusion} with a short discussion and outlook.
\end{itemize}
The paper is complemented by several technical appendices, most of which describe the details of several perturbative computations.

\section{GFF interfaces}
\label{sec:gffint}
\subsection{AdS\texorpdfstring{$_2$}{2} metric conventions}
In this paper we will be interested in interfaces arising in CFTs dual to a free scalar field in a fixed Euclidean AdS$_2$ background. We begin by reviewing the properties of the free scalar before introducing interfaces. 

Let us briefly setup conventions for the AdS$_2$ geometry. The metric in global coordinates is written
\ba
\ud s^2=L^2\,\frac{\ud r^2+ \ud \tau^2}{\cos(r)^2}\,, \qquad \mbox{with}\ r\in (-\tfrac{\pi}2,\tfrac{\pi}2)\,, \ \tau \in (-\infty,\infty)\,.
\ea
In these coordinates the boundary of AdS$_2$ is at $|r|=\frac{\pi}2$ and is composed of two infinite line segments which we should think of as the 1d cylinder $\mathds R\times S_0$ with coordinates $(\tau,\pm 1)$. Equivalently we have
\ba
\ud s^2=L^2\,\frac{\ud x^2+ \ud y^2}{y^2}\,, \qquad \mbox{with}\ y\in (0,\infty)\,, \ x \in (-\infty,\infty)\,, \label{eq:uhp}
\ea
with
\ba
x=e^{\tau} \sin(r)\,, \quad y=e^{\tau} \cos(r)\,.
\ea
Now the boundary is at $y=0$ and corresponds to the infinite line spanned by $x$. The regions $x>0$ and $x<0$ map to the lines of the 1d cylinder under $x\to (\log(|x|),\mbox{sgn}(x))$.

\subsection{Free AdS\texorpdfstring{$_2$}{2} scalar}

We consider a single free field with action
\ba
\label{eq:free-AdS2-scalar}
S_0 = \frac12\int \ud x\,\ud y\,\sqrt{g}\ \Big[(\partial \Phi)^2+m^2 L^2\, \Phi^2\Big]\:,
\ea
with
\ba
m^2 L^2:=\Delta(\Delta-1)\:.
\ea
This equation has two roots $\Delta=\Delta_{\pm}$ related by $\Delta_+=1-\Delta_-$:
\ba
\Delta_{\pm}=\frac {1\pm \sqrt{1+4 m^2 L^2}}{2}\:.
\ea
These map onto the possible allowed boundary behaviours of the bulk scalar:
\ba
\mbox{Dirichlet b.c.:}& \qquad &\Phi(x,y)\underset{y\to 0}\sim & \,y^{\Delta_+} \tilde \phi(x)\:, \\
\mbox{Neumann b.c.:}& \qquad &\Phi(x,y)\underset{y\to 0}\sim & \,y^{\Delta_-} \phi(x)\:.
\ea
In this paper we will be mostly interested in values of $m^2<0$ such that both $\Delta_{\pm}<1$. In this case both possibilities are a priori allowed and we will set
\ba
\Delta_{\tt N}\equiv \Delta_-\,, \quad \Delta_{\tt D}\equiv \Delta_+ \:.
\ea
The bulk field $\Phi$ has the mode expansion
\ba
\Phi(\tau,r)=\sum_{n=0}^{\infty} \psi_n(r)\left[e^{-E_n \tau} a_n+e^{E_n\tau} a^{\dagger}_n\right] \:,
\ea
with
\ba
\psi_n=f_{\Delta,E_n}, \qquad E_n=\Delta+n \:,
\ea
and
\ba
f_\Delta(r)&=\sqrt{\mathcal C_{\Delta}}\,\cos^{\Delta}(r)\pFq{2}{1}{\tfrac{\Delta+E_n}2,\tfrac{\Delta-E_n}2}{\tfrac 12+\Delta}{\cos^2(r)}\,, \label{solf}
\ea
being one of the solutions to the scalar equation of motion (the other being $f_{1-\Delta}$). Note that the hypergeometric function can also be expressed as a Jacobi polynomial
\ba
\psi_n(r)=\sqrt{\mathcal C_{\Delta}}\frac{n!}{\left(\Delta+\frac12\right)_n}\,(1-w^2)^\frac{\Delta}2 P_n^{\left(\Delta-\frac12,\Delta-\frac12\right)}(w)\,, \quad w:=\sin(r)\:.
\ea
The $\psi_n$ are orthogonal with respect to the inner product:
\ba
\int_{-\frac{\pi}2}^{\frac{\pi}2} \ud r\, \psi_n(r) \psi_m(r)=\delta_{n,m}\,\mathcal N_n^ 2\:,
\ea
where
\ba
\mathcal N_n^2:=\frac{n!}{ (2\Delta)_n 2 E_n}\,,\quad  \mathcal C_\Delta=\frac{\Gamma(\Delta)}{2\sqrt{\pi} \Gamma(\Delta+\frac 12)}\:.
\ea
In the quantum theory the bulk theory is dual to a generalized free field on the boundary~$\phi$ of dimension $\Delta$. With the normalisation conventions above we have
\ba
\Phi(x,y)\underset{y\to 0}\sim \sqrt{\mathcal C_{\Delta}}  y^{\Delta} \phi(x)\:,
\ea
where $\phi$ is a canonically normalised CFT primary of dimension $\Delta$. In particular,
\ba
\frac{1}{n!}\partial_x^n\phi(0) |0\rangle= a_n^{\dagger}|0\rangle\,, \qquad [a_n,a^{\dagger}_m]=\delta_{n,m} \frac{(2\Delta)_n}{n!}\:.
\ea
The bulk to bulk propagator is then
\ba
G_{BB}(\tau,\tau',r,r')&\underset{\tau>\tau'}{=}\sum_{n=0}^{\infty}\frac{e^{-E_n(\tau-\tau')}}{2 E_n}\, \frac{\psi_n(r) \psi   _n(r')}{\mathcal N_f^2\, }\\
&=\frac{\mathcal C_{\Delta}}{u^{\Delta}}\, \pFq{2}{1}{\Delta,\Delta}{2\Delta}{-\frac{4}{u}}\:, \label{bbnormal}
\ea
where
\ba
u=2\,\frac{\cosh(\tau-\tau')-\cos(r-r')}{\cos(r)\cos(r')}\:.
\ea
Pushing bulk operators to the boundary we recover conformal correlators of the boundary field $\phi$. We recall some basic properties and our conventions for such correlators in appendix~\ref{gff}.

\subsection{GFF interfaces}
We now consider interfaces between different GFF theories. We do this by imagining that $AdS_2$ space is split into two pieces separated by the geodesic $r=0$. On the left ($l$)/right ($r$) halves of the space we choose mass terms $m_{l}^2$/$m_{r}^2$ and impose boundary conditions such that on each half we have a generalised free field of dimension $\Delta_{l}$ and $\Delta_{r}$ respectively. Ultimately we will be interested in the case where $m_l=m_r$ (which still allows $\Delta_l\neq \Delta_r$), but we keep the discussion general for now.

\begin{figure}[t]
    \centering
\begin{tabular}{lr}
\begin{tikzpicture}[scale=0.8]

\def\W{4}
\def\H{5}

\fill[blue!4] (0,0) rectangle (\W,\H);
\draw[very thick] (0,0) -- (0,\H);
\draw[very thick] (\W,0) -- (\W,\H);

\draw[dashed, thick] (\W/2,0) -- (\W/2,\H);

\node at (1,2.5) {$m_l^2\,\Phi^2$};
\node at (3,2.5) {$m_r^2\,\Phi^2$};

\node[right] at (-2,2.5) {$\mathrm{GFF}_{\Delta_l}$};
\node[left]  at (\W+2,2.5) {$\mathrm{GFF}_{\Delta_r}$};

\end{tikzpicture}
&
\begin{tikzpicture}[scale=0.8]

  \draw[very thick, fill=blue!4] (0,0) circle (2);

  \draw[thick,dashed] (0,-2) -- (0,2);

  \fill (0,2) circle (2.2pt);
  \fill (0,-2) circle (2.2pt);

  \node[above] at (0,2) {$ \mu^{rl}$};
  \node[below] at (0,-2) {$\mu^{lr}$};

  \node at (-0.9,0) {$m_l^2\Phi^2$};
  \node at ( 0.9,0) {$m_r^2\Phi^2$};

  \node[left]  at (-2.15,0) {$\mathrm{GFF}_{\Delta_l}$};
  \node[right] at ( 2.15,0) {$\mathrm{GFF}_{\Delta_r}$};

\end{tikzpicture}
\end{tabular}
\caption{A free bulk scalar in AdS$_2$. The mass term is different on the left and right halves of the space split by the geodesic $r=0$. In the infinite past/future the geodesic ends on special points which act as interfaces between the CFTs living on the left and right boundaries of AdS$_2$, denoted as insertions of operators $\mu$.
\label{split}}

\end{figure}
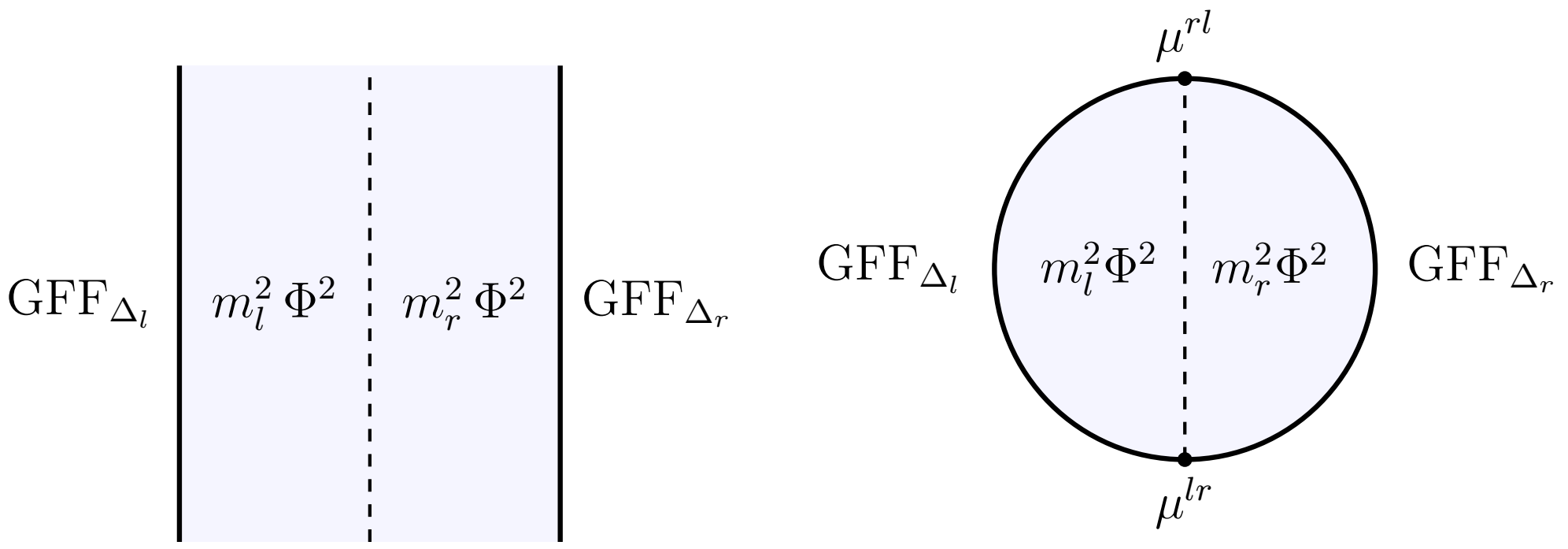

In this setup we can think that at the end of the geodesic, at $\tau=\pm \infty$, there are interfaces between the two generalized free field theories. These localized insertions we will denote as operators $\mu$. We illustrate this in figure \ref{split}.
Note that the interfaces break generically the $SL(2,R)$ isometries of AdS$_2$ down to time translations, or equivalently dilatations $x\partial_x$ on the boundary. An exception is the case $m_l^2=m_r^2$ to be addressed starting from section 3.

\subsubsection{Interface spectrum}
Perhaps the first question we can ask about the interface is what is the (renormalized) energy of the interface, what we would call $\Delta_\mu$. But to determine it is easier to start with a simpler question, which is to determine the spectrum of excitations of the interface.

To do this we must look for solutions of the free wave equation with definite energy. We have seen above these solutions generically take the form
\ba
\Phi(t,r)=e^{-E t}\, f_{\Delta,E}(r)\:,
\ea
where the scaling dimension will depend on which half of AdS$_2$ we sit.
For generic values of $E$ these functions  have a cusp at the centre of AdS$_2$. In the case without interface the dimension $\Delta$ is equal on the left and right, and the cusp disappears by tuning $E=\Delta+n$. 
In the current setup we take the solution \reef{solf} with $\Delta_l$ for $r<0$ and $\Delta_r$ for $r>0$ and we again tune $E$ so that the cusp in the total wave function vanishes. More precisely, first set
\ba
f_{E}^{lr}(r):=\theta(r) A f_{\Delta_r,E}(r)+\theta(-r) B f_{\Delta_l,E}(r)\:,
\ea
with
\ba
B= A f_{\Delta_r,E}(0)/f_{\Delta_l,E}(0)\:.
\ea
Continuity of the $r$ derivative now gives
\ba
f_{\Delta_r,E}(0^+)f_{\Delta_l,E}'(0^-)-
f_{\Delta_l,E}(0^-)f_{\Delta_r,E}'(0^+)=0\:.
\ea
This can be written as
\ba
\frac{ \Gamma \left(\Delta _l+\frac{1}{2}\right)
\Gamma \left(\Delta _r+\frac{1}{2}\right)}{\Gamma
\left(\frac{\Delta
_l-E+1}2\right) \Gamma \left(\frac
{\Delta _l+E+1}2\right) \Gamma
\left(\frac{\Delta _r-E}2\right)
\Gamma \left(\frac{\Delta
_r+E}2\right)}+(\Delta_l\leftrightarrow \Delta_r)=0\:.
\ea
This has non-trivial solutions for the $E_n$.
Since the theory is free we can provide a complete description of the interface Hilbert space via the Hamiltonian
\ba
H_{\mbox{\tiny int}}=\sum_{n=0}^\infty E_n\, b_n^{\dagger}b_n\:,
\ea
with normalised $b_n, b_n^\dagger$ ladder operators for the mode of energy $E_n$.  The spectrum in general does not have a simple closed form but we can make a few remarks.
\begin{itemize}
\item For generic $\Delta_{l,r}$ 
\ba
E_n \underset{n\to \infty}\sim E_n^{\mbox{\tiny as}}=\frac{\Delta_l+\Delta_r}2+n\:.
\ea
\item When $\Delta_l=\Delta_r+2k$ there are $k$ non-trivial algebraic values followed by $E_n=E_n^{\mbox{\tiny as}}$ exactly. 
\item For $\Delta_l\sim \Delta_r$ we have 
\ba
E_n=E_n^{\mbox{\tiny as}}+O\big[(\Delta_l-\Delta_r)^2\big]\:.
\ea
\item For $\Delta_r \gg \Delta_l$ we have with $x:=2n/\Delta_r$ fixed:
\ba
E_n&\underset{\Delta_r\to \infty}\sim \Delta_l+1+2n+\left(\frac{2}{\pi}\mbox{arccos}(x)-1\right),\\
E_n&\underset{\Delta_r\to \infty}\sim E_n^{\mbox{\tiny as}}-\frac{(-1)^n}{\pi}
\mbox{arcsin}
\left(
\tfrac{1+x-\sqrt{(x-1) (x+3)}}{1+x+\sqrt{(x-1) (x+3)}}
\sin \left[\frac{\pi}{2} 
   (\Delta_r-\Delta_l)\right]\right),
\ea
\end{itemize}
valid for $0<x<1$ and $x>1$ respectively and up to $O(\Delta_r^{-1})$ corrections.

\subsubsection{Renormalised interface energy}

Let us now compute the renormalised interface energy which we will denote as $\Delta_\mu$. To do this we compute the log of the Euclidean partition function on the cylinder with height $\beta\gg 1$, which generically is given by
\begin{equation}
\begin{aligned}
\log Z&=-\frac 12 \mbox{Tr} \log[-\nabla^2+m^2]=-\frac {\beta}2 \sum_{n=0}^\infty \int \frac{\ud E}{2\pi} \log(E^2+E_n^2)
\\
&=\frac {\beta}2 \sum_n \int\frac{ \ud E}{2\pi} \int_0^\infty \frac{\ud t}{t} \exp[-t(E^2+E_n^2)]
\\
&=-\frac{\beta}2  \sum_n E_n\:.
\end{aligned}
\end{equation}
The sum is divergent but can computed by zeta function regularization. Furthermore, we are interested in the energy of the interface relative to the GFF theories  with $\Delta_l$ or $\Delta_r$. Thus we set
\ba
\Delta_\mu^{lr}=\lim_{\beta\to \infty} -\beta^{-1}\log\frac{ Z^{\mbox{\tiny int}}}{\sqrt{Z^{l} Z^{r}}}=\frac 1 2 \sum_n E_n^{\mbox{\tiny int}}-\frac 14 \sum_n E_n^l-\frac 14 \sum_n E_n^r\:,
\ea
where each sum should be individually regulated. In particular
\ba
E_n^{l,r}= \Delta_{l,r}+n\quad \Rightarrow \quad \sum_n E_n^{l,r}:= \zeta_{\Delta_{l,r}}(-1)=-\frac 1{12}-\frac 12 \Delta_{l,r}(\Delta_{l,r}-1)\:,
\ea
and
\ba
\sum_n E_n^{\mbox{\tiny int}}:=\sum_{n=0}^\infty( E_n^{\mbox{\tiny int}}-E_{n}^{\mbox{\tiny as}})+\zeta_{\frac{\Delta_l+\Delta_r}2}(-1)\:.
\ea
This last sum cannot be done in closed form in general but it can be computed numerically. This gives
\be
\boxed{
\Delta_\mu^{lr}=\frac 12\sum_{n=0}^\infty( E_n^{\mbox{\tiny int}}-E_{n}^{\mbox{\tiny as}})+\frac 1{16}(\Delta_l-\Delta_r)^2
}
\ee
Using our previous results for the spectrum we can find
\ba
\Delta_\mu&\underset{\Delta_l\to \Delta_r}\sim \frac 1{16}(\Delta_l-\Delta_r)^2\,, \\
\Delta_\mu&\underset{\Delta_r\gg 1}\sim \left(\frac{4-\pi}{8\pi}\right)\,\Delta_r\:.
\ea

\section{The Neumann/Dirichlet interface}
\label{sec:ndint}
\subsection{Setup}
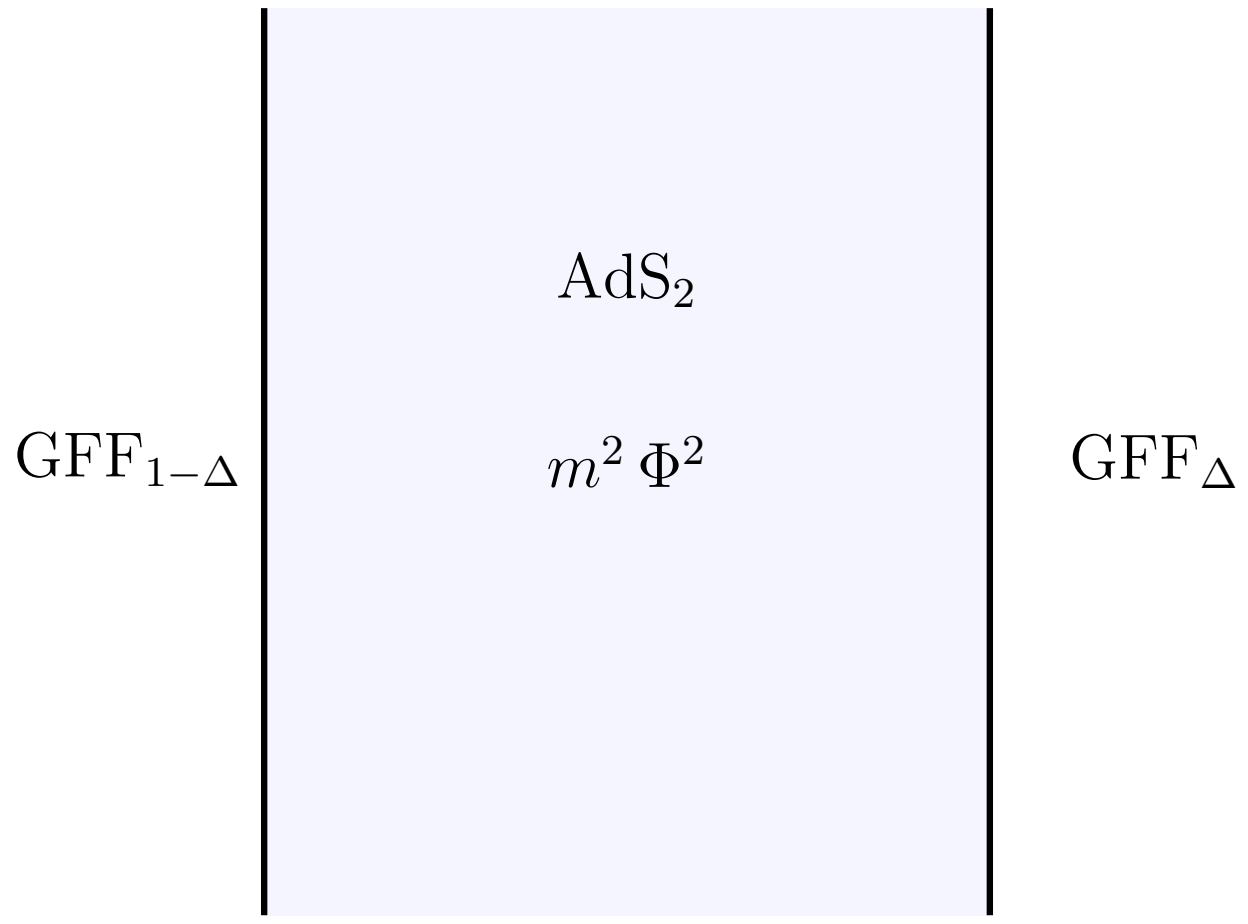
\begin{figure}[t!]
    \centering
\begin{tikzpicture}[scale=1.2]

\def\W{4}
\def\H{5}

\fill[blue!4] (0,0) rectangle (\W,\H);

\draw[very thick] (0,0) -- (0,\H);
\draw[very thick] (\W,0) -- (\W,\H);

\node at (2,3.5) {$\mathrm{AdS}_2$};
\node at (2,2.5) {$m^2\,\Phi^2$};

\node[right] at (-1.5,2.5) {$\mathrm{GFF}_{1-\Delta}$};
\node[left] at (\W+1.5,2.5) {$\mathrm{GFF}_{\Delta}$};

\end{tikzpicture}
\caption{A free bulk scalar in AdS$_2$. We choose different boundary conditions on the left and right boundaries of the space.\label{nd}}
\end{figure}
In this section we will focus on the case where $m_l^2=m_r^2$ but with distinct choices of boundary conditions. That is

\ba
\Delta_l=1-\Delta_r\equiv \Delta_{\tt D}\,, \qquad \Delta_r\equiv \Delta\equiv \Delta_{\tt N}\:.
\ea
For definiteness we take $\Delta\leq \frac 12$ so that this corresponds to an interface between Dirichlet on the left boundary of AdS$_2$ and Neumann on the right. We will also denote the Neumann GFF fields as $\phi$ and Dirichlet fields by $\tilde \phi$, which live on the right and left boundaries of AdS$_2$ respectively.

A change from Neumann to Dirichlet boundary conditions may be triggered by the addition of the relevant operator $\phi^2$ in the Neumann theory~\cite{Klebanov:1999tb} with a suitably tuned coupling. Doing so in half of the space leads to a codimension-1 interface as the one considered here. With $1-2\Delta\ll 1$ the operator is weakly relevant and a perturbative expansion can be developed as we explain in appendix \ref{app:perturbative}. 

Such interfaces have been studied in general dimension before in some detail, most notably in \cite{Melby-Thompson:2017aip} (but also  more recently in~\cite{Giombi:2024qbm}). This section acts mostly as a review of these results specialized $d=1$ CFT. There are a few new additional results which we single out, and also our interpretation of some computations is different, as we want to think of interfaces as insertions of certain operators.

One way to see this is that in this case there is no explicit breaking of the AdS$_2$ isometries in the bulk. The translation and special conformal transformation charges $P,K$ are still computed by topological operators in the bulk and are only spontaneously broken near the interfaces.
\footnote{
The spontaneous breaking means that there is a protected operator living on the interface known as the displacement~\cite{Billo:2016cpy}, which in this case has dimension $1$. In general the action of a broken charge on a conformal defect produces an integral of the displacement, but here the interface is zero dimensional and we get that the displacement is equated with the action of $P$ on the interface.
}
In particular for any state on the interface the action of $P$ is well defined and it simply shifts the energy by one unit. It follows that the interface Hilbert space must fall into representations of $SL(2,\mathds R)$. It also follows that for all intents and purposes the interface points behave as local operators which we will generically denote $\mu$, satisfying $[P,\mu]=\partial \mu$. 

We will denote $\mu^{ab}$ an operator which, on the line, implements a transition between CFT$_a$ to its left and CFT$_b$ to its right. From the perspective of the (non-local) CFT on the boundary of AdS$_2$ they act as interfaces, but from the point of view of the bulk theory they are more akin to boundary condition changing (bcc) operators in BCFT$_2$. We will accordingly sometimes denote these as interfaces or bcc operators.
\begin{figure}[t]
\centering
\begin{tikzpicture}[scale=1.1, line cap=round]

  \fill[blue!4] (-4,0) rectangle (4,2.6);

  \draw[thick, dashed] (-4,0) -- (-1.3,0);
  \draw[thick, dashed] (1.3,0) -- (4,0);

  \draw[very thick] (-1.3,0) -- (1.3,0);

  \fill (-1.3,0) circle (2pt);
  \fill (1.3,0) circle (2pt);

  \node[below=5pt] at (-1.3,0) {$\mu^{\tt N/D}$};
  \node[below=5pt] at (1.3,0) {$\mu^{\tt D/N}$};

  \node[above=5pt] at (-2.7,0) {$N$};
  \node[above=5pt] at (0,0) {$D$};
  \node[above=5pt] at (2.7,0) {$N$};

  \node at (0,1.5) {$\mathrm{AdS}_2$};
\end{tikzpicture}
\caption{Interface/bcc operators. \label{NDfig}}
\end{figure}
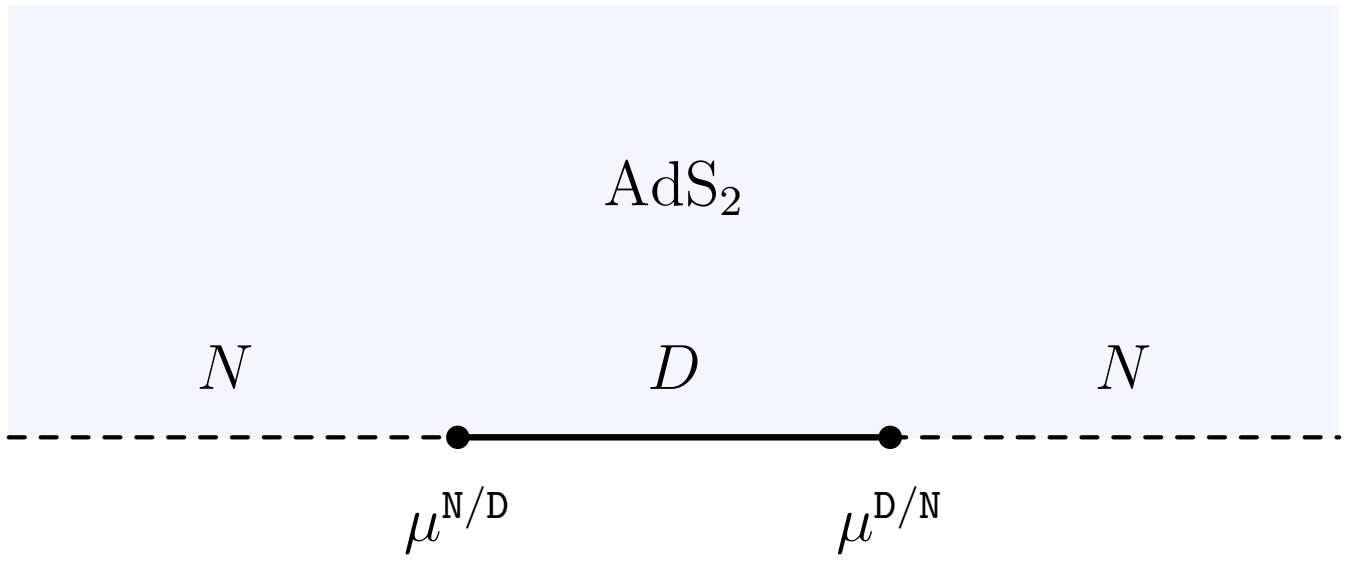

In the present setup, we have bcc operators which effect the transition from Neumann to Dirichlet boundary conditions and vice versa, see figure \ref{NDfig}. Below we will simplify notation and set
\ba
\mu\equiv \mu^{\tt N/D}\:, \qquad \tilde \mu \equiv \mu^{\tt D/N}\:.
\ea

The spectrum of interface excitations discussed in the previous section can be computed exactly for the special case of the Neumann/Dirichlet interface, namely:
\ba
E_n^{\tt N/D}=E_n^{\mbox{\tiny as}}=\frac 12+n\:.
\ea
In particular, using this spectrum we can now compute the scaling dimension of this operator~as:
\be
\boxed{
\Delta_\mu^{\tt N/D}=\frac 1{16}(1-2\Delta)^2 \:.
}
\label{eq:dmu-ND}
\ee
We will often omit the superscript to unclutter notation.
The above result means that
\ba
\langle \mu(x) \tilde \mu(0)\rangle\propto x^{-2\Delta_\mu^{\tt N/D}}\:.
\ea
This result is checked perturbatively in appendix \ref{app:perturbative}.
It is with consistent with~\cite{Melby-Thompson:2017aip} in the following sense: defining the free energy density,
\ba
\mathcal E:=F/\mbox{Vol}_{\mbox{\tiny ren}} \:,
\ea
given by the ratio of the free energy to a renormalised AdS$_{d+1}$ volume. That reference gives for general $d$
\ba
\mathcal E=\frac{1}{2(4\pi)^{\frac d2}\, \Gamma(1+\tfrac d2)}\int_0^{\frac 12-\Delta} \ud u \,u \cos(\pi u)\,\Gamma(\tfrac d2+u)\Gamma(\tfrac d2-u)\:,
\ea
which for $d=1$ gives exactly $\mathcal E=\Delta_\mu^{\tt N/D}$ as determined above.

\subsection{Eigenfunctions and bulk to bulk propagator}

In this section we will determine the bulk to bulk propagator in closed form. We begin by considering the spectrum of excitations of the $\tt N/D$ interface in more detail. 
We again express the bulk field as
\ba
\Phi(t,r)=\sum_n \psi^{\tt N/D}_n(r)\left(e^{-E_n^{\tt N/D} t} b_n+e^{E_n^{\tt N/D} t} b^\dagger_n\right).
\ea
In the special case under consideration, we may write the wavefunctions in the form
\ba
\psi_n^{\tt N/D}(r) = \sqrt{\mathcal{C}_\Delta}\frac{2^{\Delta-\frac 12+n} n!}{(\Delta+\frac12)_n} \sqrt[4]{1-w^2} \left(\frac{1-w}{1+w}\right)^{\frac{2\Delta-1}4}
P_n^{\left(\Delta-\frac12,\frac12-\Delta\right)}(w)\,, \quad w:=\sin(r)\:,
\ea
with $P_n$ Jacobi polynomials. This clearly satisfies the boundary conditions we have chosen.
The normalization then follows directly from the normalization of Jacobi polynomials.
\ba
 (\mathcal N_n^{\tt N/D})^2:=\int_{-\frac{\pi}2}^{\frac{\pi}2} \ud r\, |\psi_n^{\tt N/D}(r)|^2=
\frac{4^{n}\Gamma \left(\frac{3}{2}+n-\Delta\right)\Gamma(2\Delta)}{\Gamma\left(\Delta+\frac12+n\right)(2E_n)}\:.
\ea
We can now compute the bulk to bulk propagator:
\ba
G_{BB}^{\tt N/D}(t,t',r,r')&=\langle \mu| \Phi(t',r')\Phi(t,r) |\mu\rangle\\
&=\sum_{n=0}^\infty x^{n+\frac12}\, \frac{\psi_n^{\tt N/D}(r) \psi_n^{\tt N/D}(r')}{(1+2n)\,(\mathcal N_n^{\tt N/D})^2} \:,
\ea
where without loss of generality we chose $t<t'$ and set $x=e^{t-t'}$. 
Using a variant of Bailey's formula it is possible to obtain the closed form formula:
\ba
G_{BB}^{\tt N/D}(t,t',r,r')&= \frac{1}{2\,\Gamma\big(\Delta+\frac12\big)\Gamma\big(\frac32-\Delta\big)}
\frac{\cos ^{\Delta }(r)}{\left(1+\sin (r)\right)^{\Delta-\frac{1}{2}}}\frac{\cos ^{\Delta }(r')}{\left(1+\sin (r')\right)^{\Delta-\frac{1}{2}}} \\
&\frac{\sqrt{x}}{x+1}\ F_4\left(1,\tfrac12,\tfrac32-\Delta,\tfrac12+\Delta,\tfrac {x(1+\sin r)(1+\sin r')}{(1+x)^2},\tfrac {x(1-\sin r)(1-\sin r')}{(1+x)^2} \right),
\ea
which is a new result. $F_4$ is the fourth Appell series. As a cross-check, the above reduces to the usual bulk-to-bulk propagator \reef{bbnormal} in the special case~$\Delta=\frac 12$. 

\subsection{Boundary correlators}

Let us now discuss the results for correlators of boundary insertions. We can obtain such correlators by starting with purely bulk insertions and pushing them to the boundary. In~the present case of the $\tt N/D$ interface we can write
\ba
\Phi(t,r)&\underset{r\to -\frac \pi 2}\sim (\tfrac \pi 2+r)^{1-\Delta}\, \sqrt{\mathcal C_{1-\Delta}}\, e^{(1-\Delta) t}\, \tilde \phi(-e^{t})\:, \\
\Phi(t,r)&\underset{r\to +\frac \pi 2}\sim (\tfrac \pi 2-r)^{\Delta}\, \sqrt{\mathcal C_\Delta}\, e^{\Delta t}\, \phi(e^{t})\:,
\ea
where the constants are chosen in such a way that $\phi, \tilde \phi$ have canonical normalisation. This yields

\ba
\langle \tilde \mu| \phi(x_1) \phi(x_2)|\mu\rangle
&=\frac{1}{(x_{12})^{2 \DN }}-\frac{1}{x_1^{2\DN}}\frac{\Gamma(\tfrac 12+\DN)}{\Gamma(\tfrac 12-\DN)}
\, \pFqt{2}{1}{2\DN,\frac{1}{2}+\DN}{1+2\DN}{1-\frac{x_2}{x_1}}\,,
\\
\langle \tilde \mu| \tilde \phi(x_1) \tilde \phi(x_2)|\mu\rangle
&=\frac{1}{(x_{12})^{2 \DD }}-\frac{1}{x_1^{2\DD}}\frac{\Gamma(\tfrac 12+\DD)}{\Gamma(\tfrac 12-\DD)}
\, \pFqt{2}{1}{2\DD,\frac{1}{2}+\DD}{1+2\DD}{1-\frac{x_2}{x_1}}\,,\\
\langle \tilde \mu| \tilde \phi(x_1) \phi(x_2)|\mu\rangle&=
\left(-\frac{x_1}{x_2}\right)^{\Delta -\frac{1}{2}}
\, 
\frac{
\sqrt{\frac{1}{\Gamma (2-2 \Delta ) \Gamma (2 \Delta )}}
}{x_{21}}\:.
\label{eq:<muphiphimu>}
\ea
From these expressions it is possible to extract various pieces of CFT data (cf. appendix~\ref{gff}). For instance we have
\ba
\langle \tilde \mu|\phi(1)\phi(1-z)|\mu\rangle=z^{-2\Delta}+\sum_{n=0}^\infty \lambda^{\phi \phi}_{(\phi^2)_n} \lambda^{\mu \tilde \mu}_{(\phi^2)_n} z^{\Delta_\mu-\Delta}G^{\Delta\Delta,\Delta_\mu \Delta_\mu}_{2\Delta+2n}(z)\:,
\ea
corresponding to the exchange of double trace primaries in the $\phi \times \phi$ OPE. We find
\ba
\lambda^{\phi \phi}_{(\phi^2)_n} \lambda^{\mu \tilde \mu}_{(\phi^2)_n}=
-\frac{\cos (\pi  \Delta )(2 \Delta +2 n-1)}{2^{2 n+3} \pi\, n!(\Delta +n)}\frac{\Gamma \left(n+\Delta -\frac{1}{2}\right)^2 \Gamma \left(n+2 \Delta -\frac{1}{2}\right)}{
\Gamma (2 \Delta -1)\Gamma\left(2 n+2 \Delta -\frac{1}{2}\right)}\:. \label{opephi2}
\ea

which leads to
\ba
\label{eq:OPE-mumuphi2_n}
\left(\lambda^{\mu \tilde \mu}_{(\phi^2)_n}\right)^2=\frac{(1-2 \Delta )^2 \cos ^2(\pi  \Delta ) 4^{2 \Delta +2 n-5} (2 \Delta +2 n-1) \Gamma \left(n+\frac{1}{2}\right) \Gamma \left(n+\Delta -\frac{1}{2}\right)^4 \Gamma \left(n+2 \Delta -\frac{1}{2}\right)}{\pi ^3 n!(\Delta +n)^2 \Gamma (n+2 \Delta ) \Gamma (4 n+4 \Delta -2)}\:.
\ea
We can also do the OPE in the other channel. In this case it is useful to use a hypergeometric function identity to write
\ba
\langle \tilde \mu|\phi(1)\phi(1-z)|\mu\rangle&=\frac{\sqrt{\pi }\, 2^{1-2 \Delta }}{\Gamma(\Delta)\Gamma(\frac32-\Delta)}\,(1-z)^{\frac 12-\Delta} \pFq{2}{1}{1,\tfrac12+\Delta}{\tfrac32-\Delta}{1-z}\\
&=\sum_{n=0}^\infty \big(\lambda^{\mu\phi}_{(\mu\phi)_n}\big)^2\, (1-z)^{\Delta_\mu-\Delta} G_{\Delta_\mu+\frac 12+n}^{\Delta_\mu\Delta,\Delta\Delta_\mu}(1-z)\:.
\ea
As expected we now find operators $(\mu \phi)_n$ with dimensions $\Delta_n=\Delta_\mu+\frac 12+n$.\footnote{Note that all of $\mu\times \phi$,  $\mu \times \tilde \phi$, $\tilde \mu\times \phi$ and $\tilde \mu \times \tilde \phi$ fuse onto the same operators on the interface which we are denoting $(\mu \phi)_n$.} We find the new result
\ba
\big(\lambda^{\mu\phi}_{(\mu\phi)_n}\big)^2=\frac{(-1)^n \Gamma \left(n+\Delta +\frac{1}{2}\right)^2 \, \pFqt{3}{2}{-n,1,2
   \Delta_\mu+n}{\frac{3}{2}-\Delta ,\Delta +\frac{1}{2}}{1}}{n!\Gamma (2 \Delta )  (2\Delta_\mu+n)_n}\:, \label{eq:mufisq}
\ea
which in spite of appearances is always positive, at least if $\Delta_\mu^{\tt N/D}$ is given by \reef{eq:dmu-ND}. Note that similar results for $\lambda^{\mu \tilde \phi}_{(\mu \phi)_n}$ are obtained by sending $\Delta\to 1-\Delta$ in this expression.

Consider now the OPE data obtained from the mixed correlator. In this case we have
\ba
\langle \mu| \phi(1) \tilde\phi(\tfrac{z}{z-1})|\mu\rangle&=
\frac1{\sqrt{\Gamma (2 \Delta )\Gamma (2-2 \Delta )}}
\, \frac{z^{\Delta-\frac 12}}{(1-z)^{\Delta-\frac 32}}\\
&=\sum_{n=0}^\infty \lambda^{\mu \phi}_{(\mu \phi)_n} \lambda^{\mu \tilde \phi}_{(\mu \phi)_n} (1-z) G_{\Delta_\mu+\frac 12+n}^{(1-\Delta)\Delta_\mu,\Delta\Delta_\mu}(z)\:.
\ea
We find
\ba
\lambda^{\mu \phi}_{(\mu \phi)_n} \lambda^{\mu \tilde \phi}_{(\mu \phi)_n}
   = \frac{\Gamma \left(n+\Delta +\frac{1}{2}\right)\Gamma \left(n-\Delta +\frac{3}{2}\right) \, \pFqt{3}{2}{-n,1,2
   \Delta_\mu+n}{\frac{3}{2}-\Delta ,\Delta +\frac{1}{2}}{1}}{n!\sqrt{\Gamma (2 \Delta )\Gamma (2-2\Delta )}  (2\Delta_\mu+n)_n} \:,
\ea
consistent with the result \reef{eq:mufisq} above. One can also compute higher-point correlation functions by Wick contractions. By fusing operators together, we can then obtain correlation functions of higher-order operators. For example, we have the four-point function
\ba
\label{eq:<muphi2phi2mu>}
\frac 12\langle \tilde \mu|\phi^2(x_1)\phi^2(x_2)|\mu\rangle &= 
\frac{1}{2x_1^{2\DN}x_2^{2\DN}}\left(\frac{\Gamma(\tfrac 12+\DN)}{\Gamma(\tfrac 12-\DN)\Gamma(1+2\DN)}\right)^2 \\
&+ \left(\frac{1}{(x_{12})^{2 \DN }}-\frac{1}{x_1^{2\DN}}\frac{\Gamma(\tfrac 12+\DN)}{\Gamma(\tfrac 12-\DN)}
\, \pFqt{2}{1}{2\DN,\frac{1}{2}+\DN}{1+2\DN}{1-\frac{x_2}{x_1}}\right)^2.
\ea
with the overall $1/2$ factor so as to have canonically normalised operators $:\!\phi^2\!:/\sqrt{2}$. More generally set
\ba
P(x):=\lim_{y\to x}[ \langle \tilde \mu| \phi(y) \phi(x)|\mu\rangle-(x-y)^{-2\DN}]=-\frac{1}{x^{2\DN}} \frac{\Gamma(\tfrac 12+\DN)}{\Gamma(\tfrac 12-\DN)\Gamma(1+2\DN)}\:,
\ea
Then we have:
\begin{equation}
\boxed{
\frac{1}{n!}\langle \tilde \mu| \phi^{n}(x_1) \phi^{n}(x_2)\!|\mu \rangle = n!\sum_{k=0}^{\lfloor n/2\rfloor}\frac{P(x_1)^k P(x_2)^k}{2^{2k}(k!)^2(n-2k)!}\, (\langle \tilde \mu| \phi(x_1) \phi(x_2)|\mu\rangle)^{n-2k}\:.}
\label{defectcorrs}
\end{equation}

From this it is now easy to obtain the family of OPE coefficients:
\ba
    \lambda^{\mu \tilde \mu}_{\phi^{2n}}=
    \frac{\sqrt{(2n)!}}{2^n n!}
    \left(\frac{\Gamma\big(\Delta+\frac12\big)}{\Gamma\big(\frac12-\Delta\big)\Gamma(2\Delta+1)}\right)^n.\label{opemumun}
\ea

\subsection{An algebraic perspective on the interface Hilbert space}
Let us recall that the interface Hamiltonian is given by
\ba
H_{\tt N/D}=\sum_{n=0}^\infty E_n  b^{\dagger}_n b_n\,, \qquad E_n=\frac 12+n\:,
\ea
with $b_n, b^{\dagger}_n$ usual creation and annihilation operators. Using these operators we can construct local fields $\phi, \tilde \phi$ living on both sides of the interface,
\ba
\phi(x) &= x^{-\Delta_{\tt N}}
\sum_{n=0}^{\infty}\alpha_n(\Delta_{\tt N})
\left(x^{-E_n}b_n+x^{E_n}b_n^\dagger\right),
\qquad x>0\:,\\
\tilde{\phi}(x) &= (-x)^{-\Delta_{\tt D}}
\sum_{n=0}^{\infty}(-1)^n\alpha_n(\Delta_{\tt D})
\left((-x)^{-E_n}b_n+(-x)^{E_n}b_n^\dagger\right),
\qquad x<0\:,
\ea
with
\ba
\alpha_n(\Delta)=
\sqrt{\frac{\Gamma(n+\frac12+\Delta)}
{\Gamma(2\Delta)\Gamma(n+\frac32-\Delta)}}\:.
\ea
These are chosen so as to reproduce the two point functions \reef{eq:<muphiphimu>}.

Curiously, the interface Hamiltonian is exactly the same as the Hamiltonian for a generalized free field of dimension 1/2. Thus, using the same oscillators we could have constructed another field which we denote $\psi$, which would have had ordinary GFF correlators. Its mode decomposition corresponds to setting $\Delta_{\tt N}=\Delta_{\tt D}=1$, which gives $\alpha_n=1$.

While this may seem strange, there is no contradiction as the fields $\psi, \phi$ are not mutually local. We can make this precise by thinking of the interface conformal algebra as a deformation of that of a generalized free field of dimension $\frac 12$. For such a field we have $D=H_{\tt N/D}$ as written above as well as
\ba
P=\sum_{n=0}^\infty (n+1) b^{\dagger}_{n+1} b_n\,, \qquad K=\sum_{n=0}^\infty (n+1)  b^{\dagger}_{n} b_{n+1}\:.
\ea
They satisfy $[P,K]=2 D$ and have the expected commutation relation with the field $\psi$, but not with $\phi, \tilde \phi$. But the correct $P,K$ for those fields are easily constructed: allowing for general quadratic expressions in oscillators which raise/lower energy by one unit, one can construct:
\ba
P_{h}&=\sum_{n=0}^\infty \sqrt{(n+1)^2-h^2}b^{\dagger}_{n+1} b_n+\frac{h}2 (b_0^{\dagger})^2\,,\\
\qquad K_{h}&=\sum_{n=0}^\infty \sqrt{(n+1)^2-h^2}b^{\dagger}_{n} b_{n+1}+\frac{h}2 (b_0)^2\,,
\ea
owing to the special fact that $b^{\dagger}$ has weight 1/2 here. These operators satisfy
\ba
[P_h,K_h]=2 \left(D+\frac{h^2}4\right):=2 D_h\:.
\ea
To fix the relation of this algebra to the interface fields $\phi,\tilde \phi$, we demand
\ba
[P_{h},\phi]=\partial_x \phi\,.
\ea
Simple algebra shows that this equation holds only if
\ba
h=\frac 12-\Delta_{\tt N}\:.
\ea
We have succeeded in constructing an algebra suitable for the interface operators. But the price to pay is that the deformed dilatation operator no longer gives zero when acting on the GFF vacuum. Identifying that state with the interface we get for its energy:
\ba
\Delta_\mu= \frac{h^2}4 =\frac{( 1-2\Delta_{\tt N})^2}{16}\,.
\ea
This provides a purely algebraic derivation of our previous computation.

\section{Multi-interface Neumann/Dirichlet correlators}
\label{sec:multi-interface-correlators}
In this section we will write down a closed formula for correlators of $\tt N/D$ interface operators~$\mu$ suitable for numerical evaluation. It will be easiest in this section to work in the upper-half-plane model of AdS$_2$, \reef{eq:uhp}, where the boundary is then the real line $\mathds R$. 

\subsection{Setup and a first calculation}
We wish to compute correlation functions of $\tt N/D$ interfaces. To do so we must compute the partition function of the bulk free scalar theory with boundary conditions which may be either Neumann or Dirichlet in different regions of the AdS$_2$ boundary. 

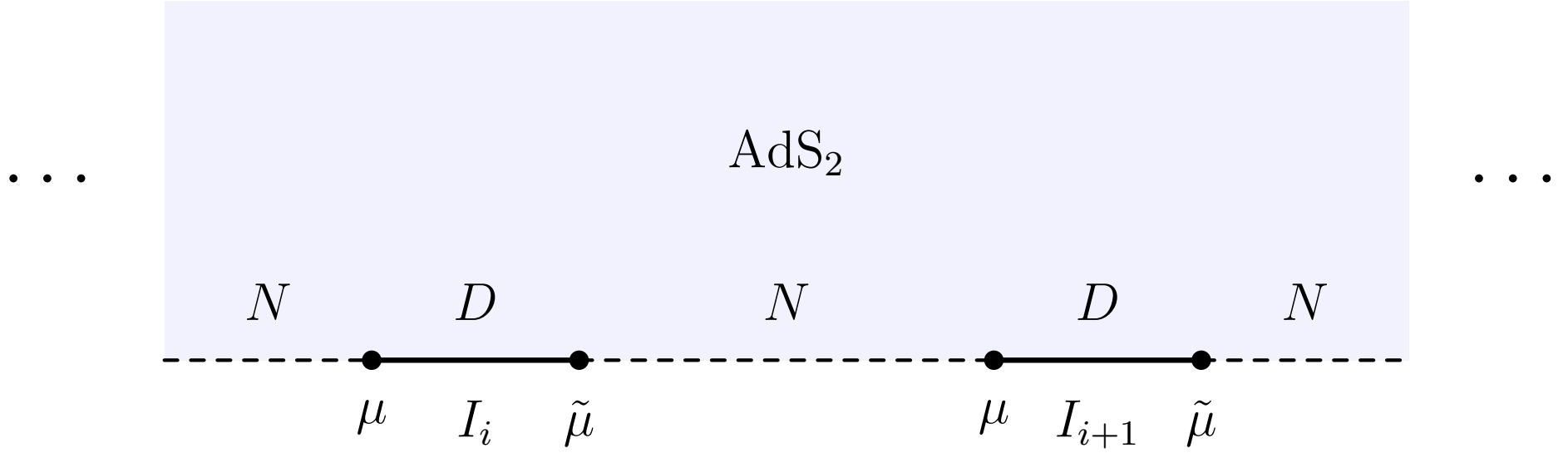
\begin{figure}[t]
\centering
\begin{tikzpicture}[scale=1.1, line cap=round]

\fill[blue!5] (-4.5,0) rectangle (4.5,2.6);

\draw[thick,dashed] (-4.5,0) -- (-3,0);
\draw[thick,dashed] (-1.5,0) -- (1.5,0);
\draw[thick,dashed] (3,0) -- (4.5,0);

\draw[very thick] (-3,0) -- (-1.5,0);
\draw[very thick] (1.5,0) -- (3,0);

\foreach \x in {-3,1.5} {
    \fill (\x,0) circle (2pt);
    \node[below=5pt] at (\x,0) {$\mu$};
}
\foreach \x in {-1.5,3} {
    \fill (\x,0) circle (2pt);
    \node[below=5pt] at (\x,0) {$\tilde\mu$};
}

\node[above=5pt] at (-3.75,0) {$N$};
\node[above=5pt] at (-2.25,0) {$D$};
\node[above=5pt] at (0,0) {$N$};
\node[above=5pt] at (2.25,0) {$D$};
\node[above=5pt] at (3.75,0) {$N$};

\node[below=5pt] at (-2.25,0) {$I_i$};
\node[below=5pt] at (2.25,0) {$I_{i+1}$};

\node at (0,1.5) {$\mathrm{AdS}_2$};

\node[font=\Large] at (-5.3,1.3) {$\cdots$};
\node[font=\Large] at (5.3,1.3) {$\cdots$};

\end{tikzpicture}
\caption{Multiple-interface correlator setup. \label{figmanyint}}
\end{figure}

For definiteness then we choose Dirichlet boundary conditions on $I\subset \mathds R$, and Neumann on the complement $\mathds R\backslash I$, with $I$ a collection of $N$ disjoint intervals, $I=\cup_{i=1}^N I_i$, see figure~\ref{figmanyint}.
The partition function for this choice of boundary conditions is denoted $Z_I$. In the case where we choose Neumann/Dirichlet boundary conditions for the full boundary we call the partition function $Z^{\tt N/D}$. 

The partition function $Z_I$ encodes the interface correlators as follows. Let us choose for instance $I=[x_0,x_1]\cup [x_2,x_3]$. This corresponds to having two Dirichlet regions. These regions are in a sense  operators $\mathcal D_{[x_0,x_1]}$ and $\mathcal D_{[x_2,x_3]}$, which we have seen can also be thought of as pairs of interface operators. We thus have
\ba
Z_I\sim \langle \mathcal D_{[x_0,x_1]} \mathcal D_{[x_2,x_3]}\rangle\:,
\ea
The partition function suffers from divergences of various kinds, and the normalisation factor is thus scheme dependent. To fix all ambiguities we define
\ba
\Big\langle \prod_{i=1}^N \mathcal D_{I_i}\Big\rangle:= \frac{\mathcal Z_{I}}{\prod_{i=1}^N\mathcal Z_{I_i}}\:,\qquad \mathcal Z_I:=Z_I/Z_{\tt N}\:. \label{intcorr}
\ea
This definition gives in particular that whenever an interval $I$ shrinks to zero size, we have $\mathcal D_I\to 1$. This means that when translated into pairs of interface operators we have
\ba
\mathcal D_{[x,y]}=(y-x)^{2\Delta_\mu}\, \mu(x)\tilde \mu(y)\:.
\ea

Let us now show how these correlators may be computed. As we have mentioned previously, a Dirichlet region in the Neumann theory may be obtained by adding a $\phi^2$ term to the action. Consider then the path integral for the Neumann GFF in the presence of the $\phi^2$ interaction, switched on along the multi-interval $I$. By introducing a Lagrange multiplier field we write this as
\ba
Z_I=\int DJ\,D\phi\, \exp\left(-S[J,\phi]\right)\:,
\ea
with
\ba
S[J,\phi]=\frac 12 \phi\cdot C^{-1}\cdot \phi-i \sqrt{2} J\cdot \phi+\frac{1}{g_0} J\cdot J\:, \qquad J(x)=\sum_{i=1}^N \chi_{I_i}(x) J^i(x)\:,
\ea
and $\chi_S(x)$ the characteristic function for set $S$.
Doing the $\phi$ path integral gives
\ba
Z_I=Z_{\tt N} \int DJ \exp[-J\cdot(\tfrac{1}{g_0}\mathds 1_I+C_I)\cdot J]\:.
\ea
Tuning to the Dirichlet fixed point sends $g_0\to \infty$. Doing the $J$ path integral it is now easy to find
\be
\label{eq:Ds-correlator}
\boxed{
\Big\langle \prod_{i=1}^N \mathcal D_{I_i}\Big\rangle =\left[ \frac{\det C_I}
{\prod_{i=1}^N \det C_{I_i}}\right]^{-\frac 12}, \qquad I=\bigcup_{i=1}^N I_i\:,
}
\ee
with $C_I$ the operator defined by
\ba
C_I[f]=g \Leftrightarrow g(y):=\int_I \frac{\ud x}{(y-x)^{2\Delta}} f(x)\:.
\ea

\subsection{Computation via bulk-to-bulk propagator}
As a cross check we will now repeat the computation in a somewhat different manner but leading to the same result. Along the way we will determine the bulk field $\Phi$ bulk to bulk propagator in the presence of an arbitrary number of interfaces. 

The idea is to compute the partition function using a trick~\cite{Kamela:1998mb}. Starting from the usual expression
\ba
\log Z_I=-\frac 12 \mbox{Tr} \log[(\Box+m^2)_I]+c\:,
\ea
for some scheme dependent constant $c$, we consider instead
\ba
\partial_{m^2} \log Z_I=-\frac 12 \mbox{Tr}\frac{1}{\Box+m^2}=-\frac 12 \int_{\mbox{\tiny AdS}} \ud X G^I_{bb}(X,X)\:,
\ea
where $G_{bb}^I$ stands for the bulk-to-bulk propagator evaluated at coincident points, which is of course divergent. But we want instead
\ba
\partial_{m^2} \log Z_I/Z^{\tt N}=-\frac 12 \int_{\mbox{\tiny AdS}} \ud X [G^I_{bb}(X,X)-G^{\tt N}_{bb}(X,X)]\:,
\ea
which does not suffer from bulk divergences but, as we will see, does diverge near the insertion points of the $\mu$ operators. These are removed by dividing by single interval partition functions as explained above.

\subsubsection*{Bulk-to-bulk propagator}
To proceed we need to determine the bulk-to-bulk propagator for the arbitrary choice of boundary conditions encoded in $I$. In the previous section we found the ${\tt N/D}$ propagator (which in the notation of this section would correspond to $I=[0,\infty]$) by first deriving definite energy wavefunctions and then writing the propagator as a sum over modes. Here we will adopt a different and simpler approach which works for any choice of interval $I$. 

The bulk-to-bulk propagator for any choice $I$ can be written as
\ba
G^I_{bb}(X,X')=G^{\tt N}_{bb}(X,X')+h^I(X,X')\,,
\ea
where $h^I(X,X')$ is a solution of the homogeneous bulk equation of
motion, constructed in such a way that the correct boundary conditions
are satisfied on $I$. A useful way to construct $h^I$ is to start from
the theory with ${\tt N}$ boundary conditions everywhere. Denote by
$\phi(x)$ the corresponding boundary field, and write
\ba
G^{\tt N}_{b\partial}(X,x)
&=&
\langle \Phi(X)\phi(x)\rangle_{\tt N}\,,
\nonumber\\
G^{\tt N}_{\partial\partial}(x,x')
&=&
\langle \phi(x)\phi(x')\rangle_{\tt N}\,.
\ea
We then seek a new bulk field $\Phi_I$, differing from $\Phi$ by a
linear combination of boundary fields supported on $I$,
\ba
\Phi_I(X)
=
\Phi(X)+\int_I dx\,f_X(x)\phi(x)\,,
\ea
such that it has vanishing correlation with the boundary field on $I$,
\ba
\langle \Phi_I(X)\phi(x')\rangle_{\tt N}=0\,,
\qquad x'\in I\,.
\ea
which is equivalent to saying that it satisfies Dirichlet rather than
Neumann boundary conditions. This condition gives the integral equation
\ba
G^{\tt N}_{b\partial}(X,x')
+
\int_I dx\,f_X(x)
G^{\tt N}_{\partial\partial}(x,x')
=0\,.
\ea
Introducing the restriction to $I$ of the Neumann boundary two-point
function,
\ba
C_I(x,x')
\equiv
G^{\tt N}_{\partial\partial}(x,x')\,,
\qquad x,x'\in I\,,
\ea
we therefore have
\ba
f_X(x)
=
-\int_I dx'\,
G^{\tt N}_{b\partial}(X,x')
C_I^{-1}(x',x)\,,
\ea
where $C_I^{-1}$ denotes the inverse of $C_I$ as an integral operator
on functions supported on $I$. The desired propagator is then simply
\ba
G^I_{bb}(X,X')
=
\langle\Phi_I(X)\Phi_I(X')\rangle_{\tt N}\,.
\ea
Using the condition above, this gives
\be
\boxed{
G^I_{bb}(X,X')
=
G^{\tt N}_{bb}(X,X')
-
\int_I dx\,dx'\,
G^{\tt N}_{b\partial}(X,x)
C_I^{-1}(x,x')
G^{\tt N}_{b\partial}(X',x')\:.
}
\ee

\subsubsection*{Computation of partition function} 
Proceeding, we have shown that
\ba
\partial_{m^2} \log Z_I/Z^{\tt N}=\frac 12 \int_I dx\,dx'\,
C_I^{-1}(x,x')
H(x,x')\:,
\ea
with
\ba
H(x,x')=\int_{\mbox{\tiny AdS}} \ud X
G^{\tt N}_{b\partial}(X,x)G^{\tt N}_{b\partial}(X,x')\:.
\ea
This integral is actually divergent from the integration region near the boundary. We can do it explicitly, regulating it by introducing a small cutoff $\epsilon$ and doing the AdS $y$ coordinate integral from $\epsilon$ to $\infty$. But a nicer way to understand the final result is to think about a related object, where we move the boundary points $x,x'$ into the bulk. In this case the above becomes
\ba
\int_{\mbox{\tiny AdS}} \ud X
G^{\tt N}_{bb}(X,X_1)G^{\tt N}_{bb}(X,X_2)=-\partial_{m^2} G_{bb}^{\tt N}(X_1,X_2)\:,
\ea
as easily follows from the equation of motion. As we now push $X_1$ and $X_2$ back to the boundary, we thus expect
\ba
H(x,x')\supset -\partial_m^2 C_I(x,x')+r C_I(x,x')\:,
\ea
where $r$ is cutoff dependent. This leads to
\ba
\partial_{m^2} \log Z_I/Z^{\tt N}=-\frac 12 \partial_m^2\mbox{Tr}[ \log C_I]+ \frac r2 \mbox{Tr}[ \mathds 1_I]\:.
\ea
Since $\mbox{Tr}[\mathds 1_{I\cup I'}]=\mbox{Tr}[\mathds 1_{I}]+\mbox{Tr}[ \mathds 1_{I'}]$ we see that cutoff dependent pieces will drop out in the computation of the interface correlators \reef{intcorr}. Putting everything together we arrive at the same result as before, namely
\be
\Big\langle \prod_{i=1}^N \mathcal D_{I_i}\Big\rangle =\left[ \frac{\det C_I}
{\prod_{i=1}^N \det C_{I_i}}\right]^{-\frac 12}, \qquad I=\bigcup_{i=1}^N I_i\:.
\ee

\subsection{The four-point correlator and the $g$-factor}

Let us consider the case where we have two Dirichlet regions, so that we are considering a~$\mu$ four-point correlator. We choose $I=[0,z]\cup[1,\infty]$. It is convenient to set:
\ba
\mathcal G(z):=(1-z)^{2\Delta_\mu} \langle \mathcal D_{[0,z]} \mathcal D_{[1,\infty]}\rangle= [z(1-z)]^{2\Delta_\mu}\langle \tilde \mu|\mu(1)\tilde \mu(z)|\mu\rangle\:,
\ea
It is then straightforward to use the expression derived above to compute the correlator numerically, essentially by discretizing the action of the integral operator $C_I$ followed by numerical diagonalisation. The power of $(1-z)$ was included to account for the expected power-law divergence in the $z\to 1$ limit. In particular a non-trivial check of the numerics is then that the final result remains finite when $z\to 1$.

\begin{figure}[H]
    \centering
    \includegraphics[height=12cm]{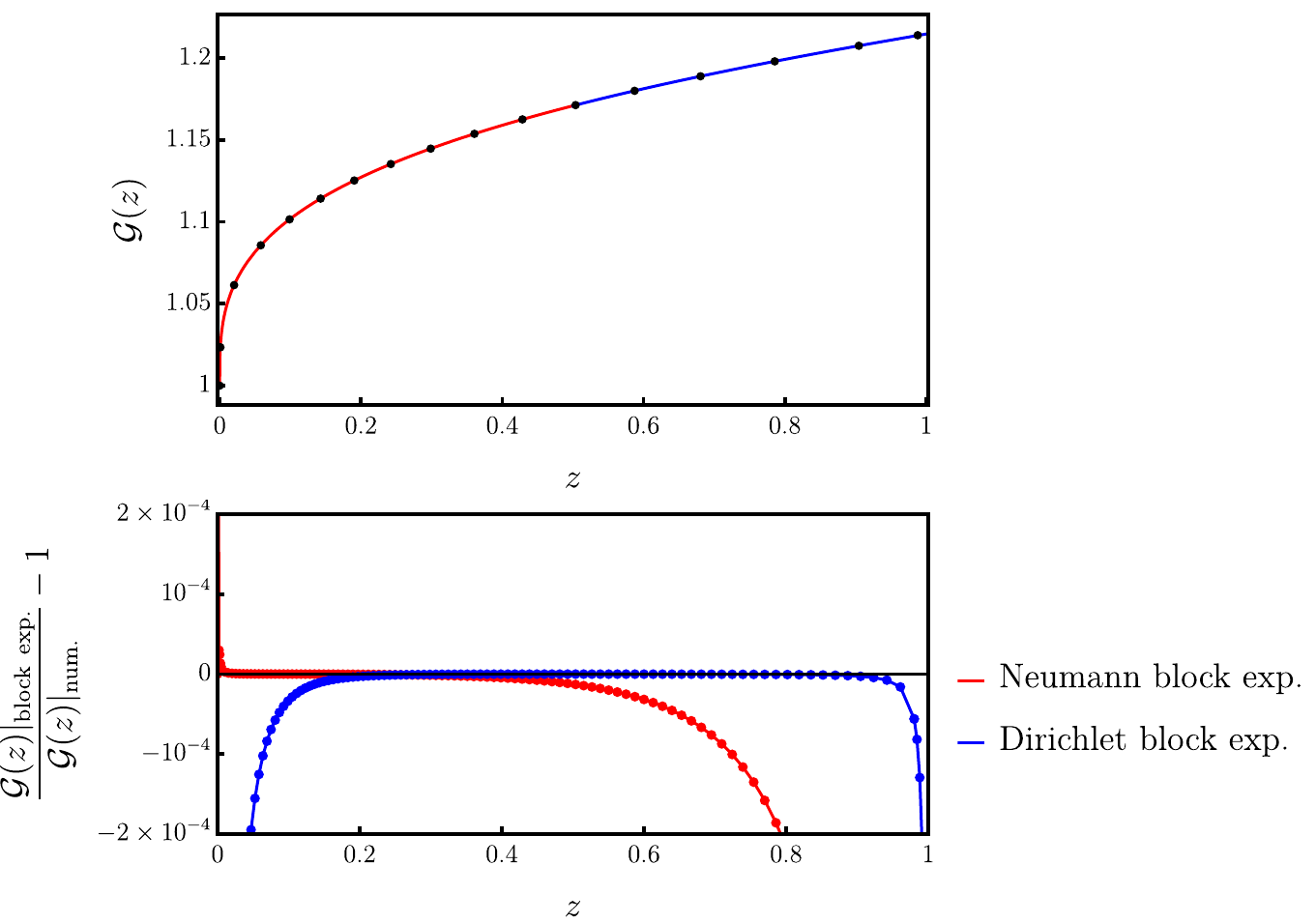}
    \caption{Numerical estimate of the four-point correlator of $\mu$ at the conformal dimension $\Delta_{\tt N}=0.15$. The blue and red lines are the perturbative block expansions using the exact spectrum of the $\tt N$ and $\tt D$ sectors.}
    \label{fig:N/D-num-G}
\end{figure}

The result is shown for a representative value of $\Delta_{\tt N}$ in figure \ref{fig:N/D-num-G}, which confirms this expectation. We also perform a comparison against the correlator written as a sum over conformal blocks, both in the $\tt N$ and the $\tt D$ sector. We~included the families of operators $(\phi^{2})_{n\leq20}$ with OPE coefficient~\eqref{eq:OPE-mumuphi2_n}, $\phi^{2n},n\leq20$ with OPE coefficient~\eqref{opemumun}, and $(\phi^4)_{n\leq20}$ with OPE coefficients extracted from block expansion of~\eqref{eq:<muphi2phi2mu>}.
As $z\to 0$ the correlator is unit by construction, but when $z\to 1$ it goes to a finite constant. To understand why, note that we have defined interface operators in such a way~that
\ba
\mu(x)\tilde \mu(0)\sim \frac{\mathds 1}{x^{2\Delta_\mu}}+\ldots
\ea
where $\mathds 1$ stands for the identity operator in the Neumann GFF theory. By construction this operator has unit vev. When we take the OPE in the other direction we will obtain now the identity in the Dirichlet theory, but nothing guarantees that it will appear with unit coefficient~\cite{Lanzetta:2025xfw}. We write
\ba
\tilde\mu(x) \mu(0)\sim \left(\frac{g_{\tt N}}{g_{\tt D}}\right)\frac{\tilde {\mathds 1}}{x^{2\Delta_\mu}}+\ldots
\ea
with $\tilde{\mathds 1}$ the Dirichlet theory identity. Consistency with the previous OPE, say in the vev of the two point function, now determines
\ba
\langle \tilde {\mathds 1}\rangle=\frac{g_{\tt D}}{g_{\tt N}}\:.
\ea
In turn this implies
\ba
\mathcal G(z)\underset{z\to 1}=\left(\frac{g_{\tt N}}{g_{\tt D}}\right)^2 \langle \tilde{\mathds 1}\rangle=\frac{g_{\tt N}}{g_{\tt D}}\:.
\ea
The quantities $\log g_{\tt N/D}$ are nothing but the the regularized free energy defined in~\cite{Bason:2025sxb}, which is essentially the difference between AdS$_2$ and S$_2$ free energies of a given QFT.\footnote{When the QFT is a CFT$_2$ then $\log g$ is identified with the Affleck–Ludwig boundary entropy \cite{Affleck:1991tk}, which measures the universal contribution of a conformal boundary condition to the partition function on a disk.} 
This is because as $z\to 1$ we go towards the pure Dirichlet theory and we expect the correlator to become $Z_{\tt D}/Z_{\tt N}\propto g_{\tt N}/g_{\tt D}$.

The GFF $g$-factors $g_N$ and $g_D$ can be exactly computed (cf.~\cite{Bason:2025sxb} table~1). One finds
\ba
\log\left(g_{\tt N}/g_{\tt D}\right)=\left(\Delta-\tfrac12\right) \log \left[2 \sin(\pi\Delta)\right]-\frac1{2\pi}\Im\left[\text{Li}_2\left(-e^{i \pi\left(1-2\Delta\right) }\right)\right].\label{gnonpert}
\ea
For $\Delta=0.15$ this gives $g_{\tt N}/g_{\tt D}\sim 1.214$, in good agreement with the numerical plot above.

As an interesting check, we can compute the four-point correlator in perturbation theory. We do this in  appendix~\ref{app:perturbative-double-inerface}. The result is
\ba
\label{eq:mumumumu}
\langle \tilde \mu| \mu(1) \tilde \mu(z)|\mu\rangle = \frac{1}{z^{2\Delta_\mu}(1-z)^{2\Delta_\mu}}\left[1+\frac{\epsilon^3}8\big(\log z \log(1-z)+2\text{Li}_2(z)\big)\right] +O(\epsilon^4)\:,
\ea
where we recall $\epsilon=1-2\Delta$. Expanding it in the Neumann and Dirichlet channels, and extract the defect entropy
\ba
\frac{g_{\tt N}}{g_{\tt D}} = 1 + \frac{\pi^2}{24}\left(1-2\Delta\right)^3+O(\epsilon^4)\:,
\ea
which matches an expansion of the non-perturbative result \reef{gnonpert}.

\section{Long range CFT interfaces}
\label{sec:lrint}
\subsection{Long range models}

Long-range models are quite generally described by free AdS$_2$ bulk theories together with boundary interactions and/or additional boundary fields. In our case we are interested in the theory of a single free bulk scalar which begins its life with Neumann boundary conditions. An equivalent description of the theory is obtained by integrating out the bulk to arrive at a boundary generalized free field,
with non-local action
\begin{equation}
    S_0 = \frac{\mathcal{N}_\phi}2\int \mathrm{d}x\,\mathrm{d}y\,\frac{\phi(x)\phi(y)}{|x-y|^{2-2\Delta_{\phi}}}\:,
\end{equation}
The coefficient $\mathcal{N}_\phi$ is fixed such that $\phi$ has unit two point function.

Interacting long range fixed points can now be 
obtained by adding interactions:
\begin{equation}
    S = S_0 + \frac{g_0}{(2n)!}\int \mathrm{d}x \,:\! \phi^{2n}(x)\!:\:.
\end{equation}
The operator $\phi^{2n}$ is weakly relevant for $\Delta_\phi=\frac{1-\epsilon}{2n}<\frac1{2n}$. Adding the interaction triggers a flow which can be tuned to end at a fixed point. The $n=1$ case corresponds to the Dirichlet theory, $n=2$ is the long-range Ising model, while $n>2$ corresponds to long range analogues of multicritical theories. These will be denoted {\tt LRI} and {\tt LR$n$I} for $n>2$.

\subsubsection{Long Range Ising}
We begin by describing some known results for the long range Ising model~\cite{Paulos:2015jfa,Behan:2017dwr,Behan:2017emf,Behan:2018hfx,Benedetti:2020rrq,Behan:2023ile,Benedetti:2024wgx,Benedetti:2025nzp}.\footnote{Recently, a preprint has appeared claiming to extend some of these computations to dramatically higher orders~\cite{Zheng:2026eti}. We will not make use of them here.}

The beta function has been computed to $O(g^4)$, here we quote the result to $O(g^3)$.
\ba
\beta(g) &= -\epsilon g + 3g^2 - 6g^3(\pi+4\log2)  + \mathcal{O}(g^4)\:.
\ea
This has a fixed point at a critical coupling
\ba
g^\star &= \frac{\epsilon}3 + \frac{2\epsilon^2(\pi+4\log2)}{9} + \mathcal{O}(\epsilon^3)\:.\label{lripert}
\ea
Perturbative computations determine
\ba
\Delta_{\phi^2}^{\tt LRI}&=\frac 12-\frac{\epsilon}6+\frac{\epsilon^2}{9}[\pi+4\log(2)]+\frac{\epsilon^3}{81} \left(\pi ^2-6\frac{\Gamma(\tfrac 14)^4}{\pi}+6\log(2)(\pi+2\log2)\right),\\
\Delta_{\phi^4}^{\tt LRI}&=1+\epsilon-\frac{2}{3} \epsilon ^2 (\pi + 4\log
   (2))+\frac{2}{27} \epsilon ^3 \left(9\pi^2-4\pi + 6\frac{\Gamma(\tfrac 14)^4}{\pi}+78\log(2)(\pi+2\log2)\right),
\ea
where we have used \reef{eq:exacti4}. For $\epsilon$ close to one we have instead:
\ba
\Delta_{\phi^2}^{\tt LRI}&=1-\sqrt{1-\epsilon}+\frac{1-\epsilon}8 \:, \\
\Delta_{\phi^4}^{\tt LRI}&=1+\sqrt{1-\epsilon}+\frac{1-\epsilon}8\:.
\ea
Using these we can write down Padé approximants which will be useful later
\begin{equation}
\begin{aligned}
\label{eq:Pade-phi2-phi4}
    \Delta_{\phi^2}^{\tt LRI} &= \frac{1 - 2.33(1-\epsilon)^{1/2} + 2.17(1-\epsilon ) - 0.808(1-\epsilon )^{3/2}}{1 - 1.33(1-\epsilon)^{1/2} + 0.712(1-\epsilon) - 0.319(1-\epsilon )^{3/2}} \:, \\
    \Delta_{\phi^4}^{\tt LRI} &= \frac{1 - 1.07(1-\epsilon)^{1/2} - 0.0372(1-\epsilon) + 0.133(1-\epsilon)^{3/2}}{1 - 2.07(1-\epsilon)^{1/2} + 1.91(1-\epsilon) - 0.809(1-\epsilon)^{3/2}} \:.
\end{aligned}
\end{equation}

\subsubsection{Long Range Multicritical models}
Let us now turn to long range multicritical models. These have not been much discussed in the literature, apart from in two dimensions under the guise of `long range minimal models' \cite{Behan:2025ydd}.

We begin by discussing the beta function. 
\ba
\beta^{\lrni}(g)=-\epsilon g+\beta_2 g^2+\beta_3 g^3+\ldots
\ea
Set $\cO= \,:\frac{\phi^{2n}}{(2n)!}:\,$. Conformal perturbation theory easily gives 
\ba
\beta_2:=\frac{C_{\cO \cO \cO}}{C_{\cO\cO}}\,, \quad \mbox{with}\quad C_{\cO\cO}:=\langle \cO|\cO\rangle\,, \quad C_{\cO\cO\cO}=\langle \cO|\cO(1)|\cO\rangle\,.
\ea
At the next order we have \cite{Komargodski:2016auf}:
\ba
\beta_3=-\frac{1}{C_{\cO \cO}}\int_0^1 \ud z\, \mathcal G_{\cO}(z)|_{\mbox{\tiny reg}}\,, \qquad \mathcal G_{\cO}(z)=\langle \cO|\cO(1)\cO(z)|\cO\rangle\,.
\ea
The regulated correlator can be chosen as
\begin{multline}
\mathcal G_{\cO}(z)|_{\mbox{\tiny reg}}:=\mathcal G_{\cO}(z)-\frac {C^2_{\cO\cO\cO}/C_{\cO\cO}}{z(1-z)}-C_{\cO\cO}^2\left(\frac{1}{z^2}+\frac{1}{(1-z)^2}+1\right)
\\
-\sum_{\Psi:0< \Delta_\Psi< 1} C_{\cO \Psi \cO}^2 \left(\frac 1{z^{2\Delta_{\cO}-\Delta_{\Psi}}}+\frac 1{(1-z)^{2\Delta_{\cO}-\Delta_{\Psi}}}-\frac{2}{1+\Delta_{\Psi}-2\Delta_{\cO}}\right).
\end{multline}
It is the first line which is more important. In the second we simply subtract power law divergent terms and add back their analytically continued integral.
We can use this result to determine the beta function for any multicritical model. We~have
\ba
C_{\cO\cO}=1/(2n)!\,, \qquad C_{\cO\cO\cO}=1/(n!)^3 \quad \Rightarrow \beta_2= \frac{(2n)!}{(n!)^3}\:.
\ea
As for the four point function it is written
\ba
\mathcal G(z)=
\sum_{\substack{a,c\geq 0\\a+c\leq 2n}}
\frac{
z^{-a/n}(1-z)^{-c/n}
}{
\bigl[a!\,c!\,(2n-a-c)!\bigr]^2
}\:.
\ea
This leads to
\begin{equation}
\boxed{
\beta_3=-(2n)!\left[
\sum_{(a,c)\in\mathcal S_n}
\frac{B\left(1-\frac{a}{n},1-\frac{c}{n}\right)}
{[a!\,c!\,(2n-a-c)!]^2}
+
2\sum_{j=0}^{n-1}
\frac{\psi(1)-\psi\left(1-\frac{j}{n}\right)}
{[n!\,j!\,(n-j)!]^2}
\right],
}
\end{equation}
where $B$ is the Beta function and
\begin{equation}
\mathcal S_n=
\left\{
(a,c)\in\mathbb Z_{\geq 0}^2:
a+c\leq 2n,\quad a\neq n,\quad c\neq n
\right\}
\setminus
\left\{(0,0),(2n,0),(0,2n)\right\}.
\end{equation}
This gives
\ba
{\tt D}:\qquad \beta_3&=0\\
{\tt LRI}:\qquad \beta_3&=-6(\pi+4\log 2)\\
{\tt LR3I}:\qquad \beta_3&\sim -126.29167599\ldots\\
{\tt LR4I}:\qquad \beta_3&\sim -193.784321088\ldots\\
\vdots\qquad & \vdots\\
\lrni:\qquad \beta_3&\underset{n\gg 1}\sim 
-\frac{3\sqrt{3}}{8\pi} B(\tfrac 13,\tfrac 13)\, \frac{3^{4n}}{n (2n)!}
\ea
where the large $n$ result comes from taking the Wick contraction terms which maximize the combinatorial factor, which happens at $a=c=2n/3$.

\subsection{Neumann/Long Range multicritical interfaces}

We will now perturbatively determine the dimension of the interface operator $\mu^{\tt N/\lrni}$ implementing a transition between the Neumann boundary condition and a long range multicritical model. The idea is to add the perturbation triggering a flow to this model only along a segment
\ba
S = S_0+\frac{g_0}{(2n)!} \int_0^T \ud x\, \phi^{2n}(x)\:,
\ea
After tuning the coupling to the fixed point, the scaling dimension can be read off from the partition function,
\ba
\Delta_\mu^{{\tt N}/\lrni} =-\frac 12 \frac{\partial\log Z(T)}{\partial \log T}\:.
\ea
Here we will perform the computation to $O(\epsilon^3)$. To this order we have
\ba
\log Z(T)=\frac{g_0^2}{2}\int_{[0,T]^2} \ud x \ud y \langle \cO(x)\cO(y)\rangle-\frac{g_0^3}{3!}\,\int_{[0,T]^3} \ud x \ud y \ud z \langle \cO(x)\cO(y)\cO(z)\rangle+O(g_0^4)\:,
\ea
where $\cO=\,:\frac{\phi^{2n}}{(2n)!}:$ as before. The integrals take the form described in appendix \ref{app:perturbative-dmu}, cf. equation \reef{integrals}, and can be done straightforwardly. The result is
\ba
\log Z(T)=-\frac{g_0^2 T^{2\epsilon}\, C_{\cO\cO}}{2\epsilon(1-2\epsilon)}+\frac{g_0^3 T^{3\epsilon} C_{\cO\cO\cO}}{3\epsilon(1-3\epsilon)}\,\frac{\Gamma(\epsilon)^2}{\Gamma(2\epsilon)}+O(g_0^4)\:.
\ea
Expressing $g_0$ in terms of the renormalised coupling and tuning to the fixed point we find
\be
\boxed{
\Delta_\mu^{{\tt N}/\lrni}=\frac{(C_{\cO\cO})^3 }{2 (C_{\cO\cO\cO})^2}\,\epsilon^2-\frac{(C_{\cO\cO})^5 \beta_3}{(C_{\cO\cO\cO})^4}\, \epsilon^3+O(\epsilon^4)\,, \qquad \epsilon=1-2n \Delta\:.
}
\ee
This gives in particular
\ba
\Delta_\mu^{\tt N/D}&=\frac{(1-2\Delta)^2}{16}+O\big((1-2\Delta)^4\big)\:,
\ea
consistent with our previous computation.

For the long range Ising model it is actually possible to push the computation further to $O(\epsilon^4)$, thanks to knowledge of the beta function to order $g^4$. 
We describe this computation in appendix~\ref{app:dnlri}. The result, consistent with the one derived above, is:
\boxeda{
\Delta_\mu^{\tt N/LRI}={}&\frac{\eps^2}{432}+\frac{\pi+4\log2}{324}\eps^3\\
&+\frac{\eps^4}{11664}\left[15\pi^2+132\pi\log2+300\log^22-8
-24\pi\frac{\G(\tfrac14)^2}{\G(\tfrac34)^2}\right]+O(\eps^5)\:.
\label{eq:dmu-NLRI}
}

\subsection{Dirichlet/Long Range multicritical interfaces}

We now turn to interfaces between the long range multicritical models and the Dirichlet theory. The idea now is to first start with the $\tt N/D$ interface with GFF of dimension $\Delta=\frac{1-\epsilon}{2n}$ and again deform the Neumann side by the $\phi^{2n}/(2n)!$  leading to $\lrni$. The Euclidean partition function is then 
\ba
\langle \mu(T) e^{-g_0 \int_0^T \ud x\, \cO} \tilde \mu(0)\rangle &\sim  \langle \mu(T) \tilde \mu(0)\rangle  - g_0 \int_0^T \ud x\,\langle \mu(T) \cO(x) \tilde \mu(0)\rangle \\
&= T^{-2\Delta_\mu^{\tt N/D}}(1- g_0 \lambda^{\mu\tilde \mu}_{\cO}\int_0^T \frac{\ud x}{T^{-2n\Delta}\big(x(T-x)\big)^{2n\Delta}} \\
&= T^{-2\Delta_\mu^{\tt N/D}}\left[1- g_0 \lambda^{\mu\tilde \mu}_{\cO}\frac{\Gamma(\epsilon)^2}{\Gamma(2\epsilon)}T^{\epsilon} \right].
\ea
Tuning to the fixed point we find
\ba
\Delta_\mu^{{\tt D}/\lrni}=\Delta_\mu^{\tt N/D}+\frac{C_{\cO\cO}}{C_{\cO\cO\cO}} \lambda^{\mu \tilde \mu}_{\cO} \epsilon\:.
\ea

Using \reef{opemumun} this becomes
\boxeda{
\Delta_\mu^{{\tt D}/\lrni}=\frac{(n-1)^2}{16 n^2}+\epsilon  \left[\frac{ (n!)^2
   }{2^n (2 n)!}\, \left(\frac{\Gamma \left(\frac{n+1}{2 n}\right)}{\Gamma
   \left(1+\frac{1}{n}\right) \Gamma \left(\frac{n-1}{2
   n}\right)}\right)^n+\frac{n-1}{8
   n^2}\right]+O(\epsilon ^2)\:.
}
As a cross check this vanishes when $n=1$. 

For the {\tt LRI} model, i.e. $n=2$, it is possible to push the computation to the next order. It~involves a regulated integral of the connected two point function of $\phi^4$ in the presence of the defect, which is hard to perform analytically but can be done numerically straightforwardly. We provide details of this computation in appendix \ref{sec:deltaDlri}. The result is
\be
\boxed{
\label{eq:dmu-DLRI}
\Delta_\mu^{\tt D/LRI} = \frac{1}{64} + \epsilon \left(\frac1{32}+\frac{\pi}{3}\,\frac1{\Gamma\big(\frac14\big)^4}\right) + \epsilon^2\, 0.0504496\ldots +O(\epsilon^3)\:.
}
\ee

\subsection{Neumann/Dirichlet/Long Range Ising three point function}
\begin{figure}[t]
\centering
\begin{tikzpicture}[scale=1.0, line cap=round, line join=round]

  \begin{scope}[shift={(0,-1)},scale=1.1]
    \fill[blue!4] (-4,0) rectangle (4,2.6);

    \draw[very thick] (-4,0) -- (-1.2,0);
    \draw[very thick, dash pattern=on 5pt off 2pt on 1.2pt off 2pt]
      (-1.2,0) -- (1.2,0);
    \draw[thick, dashed] (1.2,0) -- (4,0);

    \fill (-1.2,0) circle (2pt);
    \fill (1.2,0) circle (2pt);

    \fill (4,0) circle (2pt);

    \node[above=5pt] at (-2.6,0) {$\tt D$};
    \node[above=5pt] at (0,0) {$\tt LRI$};
    \node[above=5pt] at (2.8,0) {$\tt N$};

    \node at (0,3) {$Z^{\tt D/LRI/N}$};
  \end{scope}

  \begin{scope}[shift={(9,1.35)}, scale=0.72]
    \fill[blue!4] (-3,0) rectangle (3,2.2);

    \draw[very thick] (-3,0) -- (-0.9,0);
    \draw[very thick, dash pattern=on 5pt off 2pt on 1.2pt off 2pt]
      (-0.9,0) -- (0.9,0);
    \draw[very thick] (0.9,0) -- (3,0);

    \fill (-0.9,0) circle (2pt);
    \fill (0.9,0) circle (2pt);

    \node[above=4pt] at (-2.0,0) {$\tt D$};
    \node[above=4pt] at (0,0) {$\tt LRI$};
    \node[above=4pt] at (2.0,0) {$\tt D$};

    \node at (0,2.6) {$Z^{\tt D/LRI}$};
  \end{scope}

  \begin{scope}[shift={(9,-1.85)}, scale=0.72]
    \fill[blue!4] (-3,0) rectangle (3,2.2);

    \draw[thick, dashed] (-3,0) -- (-0.9,0);
    \draw[very thick, dash pattern=on 5pt off 2pt on 1.2pt off 2pt]
      (-0.9,0) -- (0.9,0); 
    \draw[thick, dashed] (0.9,0) -- (3,0); 

    \fill (-0.9,0) circle (2pt);
    \fill (0.9,0) circle (2pt);

    \node[above=4pt] at (-2.0,0) {$\tt N$};
    \node[above=4pt] at (0,0) {$\tt{LRI}$};
    \node[above=4pt] at (2.0,0) {$\tt N$};

    \node at (0,2.6) {$Z^{\tt N/LRI}$};
  \end{scope}

\end{tikzpicture}
\caption{Partition functions entering the normalized three-point function. 
The main object is \(Z^{\tt D/LRI/N}\), while the two-point normalization factors are 
\(Z^{\tt D/LRI}\) and \(Z^{\tt N/LRI}\). Solid, dashed, and dash-dotted segments denote 
Dirichlet, Neumann, and LRI boundary conditions, respectively. Interfaces are inserted at $0,1$ and additionally at $\infty$ in the leftmost diagram.}
\label{threepoint}
\end{figure}
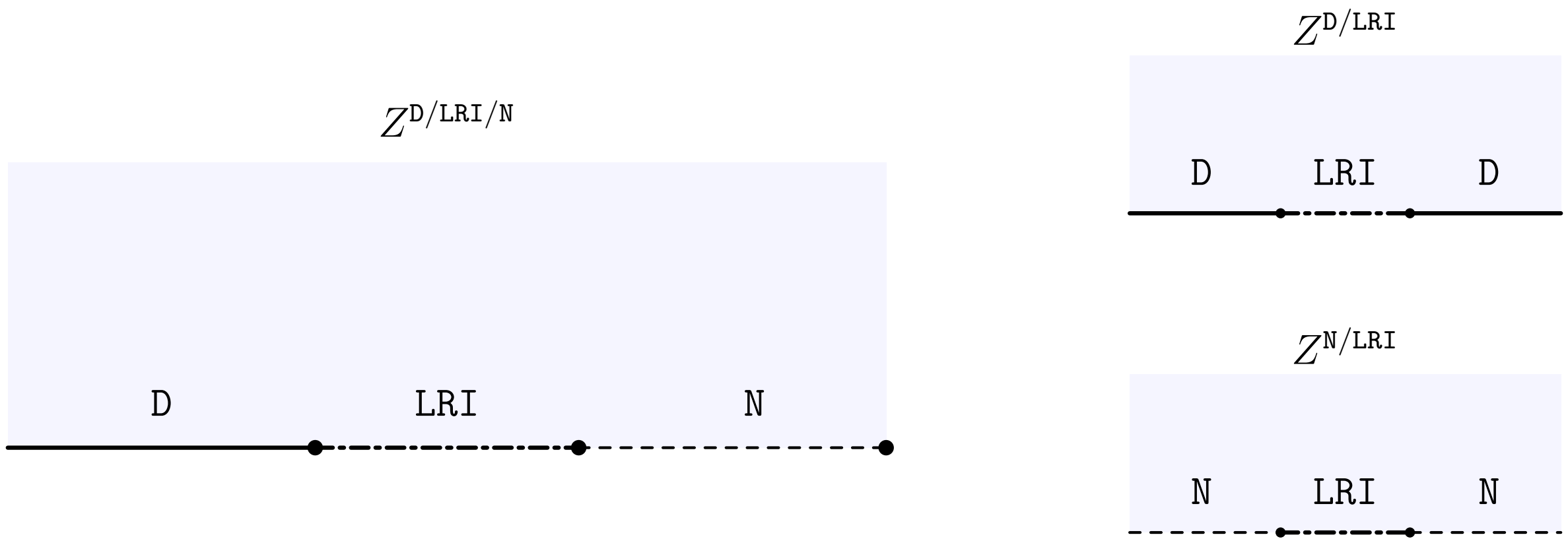

As we've seen, in the range $\Delta<1/4$ the Neumann GFF theory can be deformed both by $\phi^2$ and $\phi^4$, leading to the Dirichlet GFF and $\tt LRI$ fixed points. Having looked at interfaces between each pair in these three phases, we will now consider a three point function involving all of them, see figure \ref{threepoint}.

\noindent
The object we want to compute is
\ba
C_{\tt D/LRI/N}:=\langle \mu^{\tt N/D}| \mu^{\tt LRI/N}(1)|\mu^{\tt D/LRI}\rangle\:,
\ea
where all operator insertions should be normalised to have unit two point function.
Let us set $V:=\exp(-\frac{g_0}{4!} \int_0^1 \phi^4)$ and set
\ba
Z^{\tt D/LRI/N}&:=\langle \mu^{\tt N/D}(\infty)\, V \,\mu^{\tt D/N}(0)\rangle\:, \\
Z^{\tt D/LRI}&:=\langle \mu^{\tt N/D}(1) V  \mu^{\tt N/D}(0)\rangle\:, \\
Z^{\tt N/LRI}&:=\langle  V  \rangle\:.
\ea
Then
\ba
C_{\tt D/LRI/N}=\frac{Z^{\tt D/LRI/N}}{\left(Z^{\tt D/LRI}Z^{\tt N/LRI}\right)^{\frac 12}}\:.
\ea
Let us first compute $Z^{\tt N/LRI}$. We have
\ba
\log (Z^{\tt N/LRI})^{-\frac 12}=-\frac 12\frac{g_0^2}{4!}\, \int_{[0,1]^2} \frac{\ud x\, \ud y }{(x-y)^{2(1-\epsilon)}}=\frac{g_0^2}{48}\left(\frac{1}{2\epsilon}+1\right).
\ea
As for the other two factors, it is easier to consider them together. At order $g_0$ we have
\ba
\log\left(\frac{Z^{\tt D/LRI/N}}{(Z^{\tt D/LRI})^{\frac 12}}\right)\bigg|_{O(g_0)}&=-\,\frac{g_0\lambda^{\mu\tilde \mu}_{\phi^4}}{\sqrt{24}} \int_0^1 \ud x\left(\frac{1}{x^{1-\epsilon}}-\frac 12\frac 1{[x(1-x)]^{1-\epsilon}}\right)\\
&=-\frac{g_0 \epsilon}{6}\,\frac{\pi^3}{\Gamma(\tfrac 14)^4}+O(\epsilon^2)\:,
\ea
where we used equation \reef{opemumun} for the OPE coefficient $\lambda^{\mu\tilde \mu}_{\phi^4}$.
At order $g_0^2$ we get instead an integral over the connected two point function:
\ba
\log\left(\frac{Z^{\tt D/LRI/N}}{(Z^{\tt D/LRI})^{\frac 12}}\right)\bigg|_{O(g_0^2)}=\frac{g_0^2}{4!} J_2(\epsilon)\:,
\ea
with
\begin{equation}
    J_2(\epsilon)=\frac{1}{4!}\int_0^1\ud x\int_0^x \ud y \left[\langle \mu(\infty) \phi^4(x)\phi^4(y) \mu(0)\rangle_c-\frac 12\langle \mu(1) \phi^4(x)\phi^4(y) \mu(0)\rangle_c\right].
\end{equation}
Using the same tricks as in appendix \ref{sec:deltaDlri}, we can write this as
\ba
J_2(\epsilon)=\frac 1{4!}\int_0^1 \ud z\left\{\left(\frac 1{2\epsilon}-  \frac{\Gamma (2 \epsilon )^2}{2\Gamma(4\epsilon)} \pFq{2}{1}{2\epsilon,2\epsilon}{4\epsilon}{1-z}\right)\langle \tilde \mu| \phi^4(1)\phi^4(z)|\mu\rangle\right\},
\ea
where the two point function in the integrand is computed by \reef{defectcorrs}.
Expanding the integrand we find a term $-\frac 12(1-z)^{-1+2\epsilon}$ which is responsible for a simple $\epsilon$ pole. The finite remainder can be determined numerically (after analytic continuation in $\epsilon$):
\ba
J_2(\epsilon)=-\frac 1{4 \epsilon}+j_2+O(\epsilon)\,, \qquad j_2\sim 5.2006539\ldots
\ea
All poles cancel as they should in the full expression for $C_{\tt D/LRI/N}$. Using $g_0=g+3 g^2/\epsilon+O(g^3)$ as well as expression \reef{opemumun} for $\lambda^{\mu\tilde \mu}_{\phi^4}$, we get at the fixed point value $g_*=\epsilon/3$:
\be
\boxed{
C_{\tt D/LRI/N}=1+\frac{\epsilon^2}{432}  \left(2 j_2+1-\frac{48 \pi ^3}{\Gamma \left(\frac{1}{4}\right)^4}\right)=1+\epsilon^2\, 0.00645384194\ldots +O(\epsilon^3)\:.
}
\ee

\subsection{Computing $g$-factors}

In this section we explain how ratios of $g$-factors can be computed in perturbation theory using the formalism of~\cite{Bason:2025sxb}. We start with some 1d CFT living on the boundary of AdS$_2$ which we take to be a circle. We then add a relevant operator to the action,
\ba
S_{\mbox{\tiny int}}=\frac{\lambda}{2\pi} \int_0^{2\pi}\ud \theta\, \mathcal O(\theta)\,, \qquad \Delta_{\cO}=1-\epsilon\,.
\ea
In our application below we will take $\epsilon$ to be small. We assume that this triggers a flow to an IR fixed point. Following \cite{Bason:2025sxb} (see also~\cite{Friedan_2004} for closely related work for defects in 2d CFT) we have 
\ba
\log(g_{\mbox{\tiny UV}}/g_{\mbox{\tiny IR}})=-\int_0^{\lambda_*} \ud \lambda\, \beta(\lambda) \mathcal M_{\cO}(\lambda)\:, \label{formulag}
\ea
with $\lambda_*$ the critical coupling and
\ba
\mathcal M_{\cO}(\lambda)=\int_0^{2\pi} \frac{\ud \theta_1}{2\pi} \frac{\ud \theta_2}{2\pi}\,(1-\cos \theta_{12})\, \langle \cO(\theta_1)\cO(\theta_2)\rangle_{\mbox{\tiny int}}\:,
\ea
where the subscript indicates that two point function is evaluated in the deformed CFT. In~particular the $\cO$ stand for the renormalised operators in the interacting theory. 
Let us assume the flow to be short and work perturbatively in $\lambda$. We write the $\beta$ function as
\ba
\beta(\lambda)=-\epsilon \lambda+b\,\lambda^2+c \lambda^3+O(\lambda^4)\:,
\ea
with $b=\beta_2/(2\pi)$, $c=\beta_3/(2\pi)^2$.
The metric $\mathcal M_{\cO}(\lambda)$ is expanded as
\ba
\mathcal M_{\cO}(\lambda)=\underbrace{(M^{(0)}_0+\epsilon M^{(0)}_1+\ldots)}_{\mathcal M_{\cO}^{(0)}}+\lambda \underbrace{(M^{(1)}_0+\ldots)}_{\mathcal M_{\cO}^{(1)}}+\ldots
\ea
Using formula \reef{formulag} we find
\ba
\log(g_{\mbox{\tiny UV}}/g_{\mbox{\tiny IR}})=\frac{M^{(0)}_0}{6 b^2}\, \epsilon^3+\epsilon^4\left[\frac{M^{(0)}_1}{6b^2}-\frac{c\, M_0^{(0)}}{4 b^4}\, +\frac{M^{(1)}_0}{12 b^3}\right]+O(\epsilon^5)\:.
\ea
We compute first $M^{(0)}:=\mathcal M_{\cO}(0)$. Let us set in the UV theory
\ba
\langle \cO_0(\theta_1)\cO_0(\theta_2)\rangle&=\frac{C_{\cO\cO}}{d_{12}^{2(1-\epsilon)}}\:, \\
\langle \cO_0(\theta_1)\cO_0(\theta_2)\cO_0(\theta_3)\rangle&=\frac{C_{\cO\cO\cO}}{\prod_{i<j}^3 d_{ij}^{(1-\epsilon)}}\:, \\
\ea
with
\ba
d_{ij}:=2 \sin\frac{\theta_{ij}}2\:.
\ea
We denote in the interacting theory the renormalised operator by $\cO_r$, and the UV operator by $\cO_0$. We have
\ba
\cO_r=(1+\frac{2b \lambda}{\epsilon}+\ldots) \cO_0\:,
\ea
and hence
\ba
\mathcal M_\cO^{(0)}&=\int_0^{2\pi} \frac{\ud \theta_1}{2\pi} \frac{\ud \theta_2}{2\pi}\,(1-\cos \theta_{12})\, \langle \cO_0(\theta_1)\cO_0(\theta_2)\rangle\:,\\
\mathcal M_\cO^{(1)}&=\frac{4b}{\epsilon} \mathcal M_{\cO}^{(0)}- 2\pi \int_0^{2\pi} \prod_{i=1}^3\frac{\ud \theta_i}{2\pi} \,(1-\cos \theta_{12})\,\langle \cO_0(\theta_1)\cO_0(\theta_2)\cO_0(\theta_3)\rangle\:.
\ea
The integral in the first line can be done straightforwardly. The one second line can also be done exactly, as it is related to Dyson's circle partition integrals, see e.g. \cite{Forrester2010}. One finds
\ba
\mathcal M_{\cO}^{(0)}&=\frac{C_{\cO\cO}}2(1+O(\epsilon^2))\:,\\
\mathcal M_{\cO}^{(1)}&=O(\epsilon^2)\:.
\ea
Altogether we find
\ba
\log(g_{\mbox{\tiny UV}}/g_{\mbox{\tiny IR}})=\frac{\pi ^2 C_{\cO\cO} \epsilon ^3}{3 \beta_2^2}-\frac{2 \pi ^4 \beta_3 C_{\cO\cO} \epsilon
   ^4}{\beta_2^4}+O(\epsilon^5)\:.
\ea
As a check, consider the case where the UV theory is a Neumann GFF field. In this case the exact result~\eqref{gnonpert} expands as
\ba
\log\left(g_{\tt N}/g_{\tt D}\right)\sim \frac{\pi^2}{24} \epsilon^3+O(\epsilon^5)\,, \quad \epsilon=1-2\Delta\:,
\ea
in agreement with our perturbative computation, for which $c=0$ and $b=\frac{8}{2\pi}$.

Now consider instead the case where the flow is from Neumann GFF to the Long Range Ising model. Using the result for the beta function given in equation~\reef{lripert} (remembering that in this section we have $\lambda=2\pi g$), we find
\ba
\log\left(g_{\tt N}/g_{\tt LRI}\right)\sim \frac{\pi^2}{648} \epsilon^3\left[1+(\pi+ \log 16)\epsilon\right]+O(\epsilon^5)\:.\label{eq:gnglripert}
\ea

\section{Bootstrapping interface correlators}
\label{sec:bootstrap}
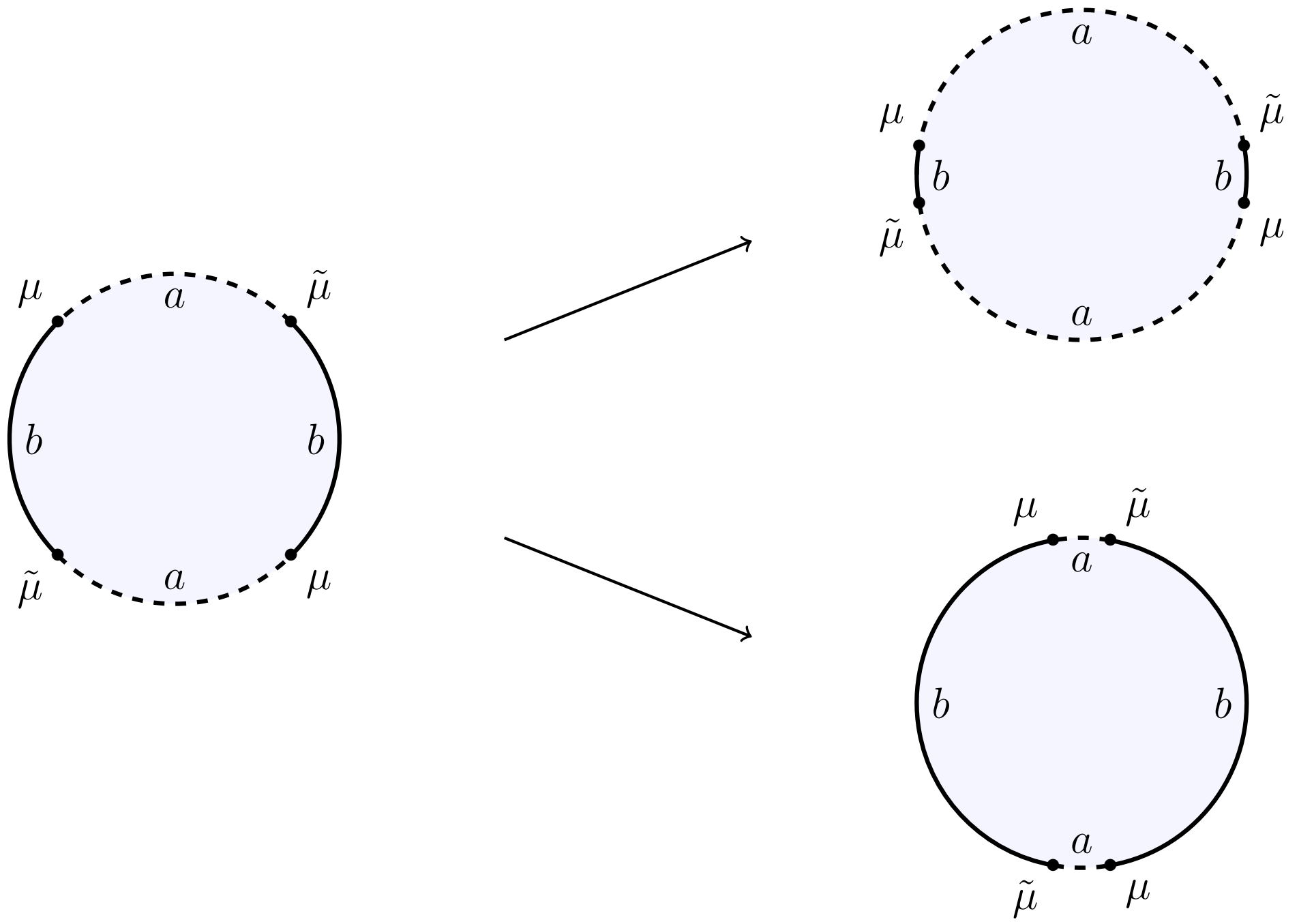
\begin{figure}[t!]
\centering
\begin{tikzpicture}[scale=1.6]
\begin{scope}
    \fill [fill=blue!4] (0,0) circle (1);
    \node (D1) at (90:1){};
    \node (N1) at (180:1){};
    \node (D2) at (270:1){};
    \node (N2) at (0:1){};
    \node (mu1) at (45:1){{\scriptsize$\bullet$}};
    \node (mu2) at (135:1){{\scriptsize$\bullet$}};
    \node (mu3) at (225:1){{\scriptsize$\bullet$}};
    \node (mu4) at (315:1){{\scriptsize$\bullet$}};

    \draw [very thick, dashed] (mu1) arc (45:135:1);
    \draw [very thick] (mu2) arc (135:225:1);
    \draw [very thick, dashed] (mu3) arc (225:315:1);
    \draw [very thick] (mu4) arc (315:405:1);
    
    \draw (D1) node [below]{$a$};        
    \draw (N1) node [right]{$b$}; 
    \draw (D2) node [above]{$a$};        
    \draw (N2) node [left]{$b$}; 
    \draw (mu1) node [above right]{$\tilde \mu$};
    \draw (mu2) node [above left]{$\mu$}; 
    \draw (mu3) node [below left]{$\tilde \mu$};
    \draw (mu4) node [below right]{$\mu$}; 
\end{scope}

\begin{scope}[xshift=5.5cm,yshift=1.6cm]
    \fill [fill=blue!4] (0,0) circle (1);
    \node (D1) at (90:1){};
    \node (N1) at (180:1){};
    \node (D2) at (270:1){};
    \node (N2) at (0:1){};
    \node (mu1) at (170:1){{\scriptsize$\bullet$}};
    \node (mu2) at (190:1){{\scriptsize$\bullet$}};
    \node (mu3) at (350:1){{\scriptsize$\bullet$}};
    \node (mu4) at (10:1){{\scriptsize$\bullet$}};

    \draw [very thick] (mu1) arc (170:190:1);
    \draw [very thick, dashed] (mu2) arc (190:350:1);
    \draw [very thick] (mu3) arc (350:370:1);
    \draw [very thick, dashed] (mu4) arc (10:180:1);
    
    \draw (D1) node [below]{$a$};        
    \draw (N1) node [right]{$b$}; 
    \draw (D2) node [above]{$a$};        
    \draw (N2) node [left]{$b$}; 
    \draw (mu1) node [above left]{$\mu $};
    \draw (mu2) node [below left]{$\tilde \mu$}; 
    \draw (mu3) node [below right]{$\mu$};
    \draw (mu4) node [above right]{$\tilde \mu$}; 
\end{scope}

\begin{scope}[xshift=5.5cm,yshift=-1.6cm]
    \fill [fill=blue!4] (0,0) circle (1);
    \node (D1) at (90:1){};
    \node (N1) at (180:1){};
    \node (D2) at (270:1){};
    \node (N2) at (0:1){};
    \node (mu1) at (100:1){{\scriptsize$\bullet$}};
    \node (mu2) at (260:1){{\scriptsize$\bullet$}};
    \node (mu3) at (280:1){{\scriptsize$\bullet$}};
    \node (mu4) at (80:1){{\scriptsize$\bullet$}};

    \draw [very thick] (mu1) arc (100:260:1);
    \draw [very thick, dashed] (mu2) arc (260:280:1);
    \draw [very thick] (mu3) arc (280:440:1);
    \draw [very thick, dashed] (mu4) arc (80:100:1);
    
    \draw (D1) node [below]{$a$};        
    \draw (N1) node [right]{$b$}; 
    \draw (D2) node [above]{$a$};        
    \draw (N2) node [left]{$b$}; 
    \draw (mu1) node [above left]{$\mu $};
    \draw (mu2) node [below left]{$\tilde \mu$}; 
    \draw (mu3) node [below right]{$\mu$};
    \draw (mu4) node [above right]{$\tilde \mu$}; 
\end{scope}  

\draw[->,thick] (2,0.6) -- (3.5,1.2);
\draw[->,thick] (2,-0.6) -- (3.5,-1.2);

\end{tikzpicture}
\caption{A two-region correlator and its two OPE limits}
\label{fig:circles}
\end{figure}

In this last section we will use conformal bootstrap techniques to constrain four point interface correlators between two CFTs/boundary conditions labeled $a,b$, as shown schematically in figure \ref{fig:circles}. Previous related works include~\cite{Lanzetta:2025xfw,Maeda:2026tuw,Furuta:2026xnz}.
In this section we denote the corresponding interface operators as $\mu^{ab}\equiv \mu$ and $\mu^{ba}\equiv \tilde \mu$. Note that $\mu \longleftrightarrow \tilde \mu$ under parity (assuming both $a,b$ boundary conditions are parity preserving). The correlator is written
\ba
\mathcal G(z):=\langle \mu^{ba}|\, \mu^{ab}(1)\mu^{ba}(z)|\mu^{ab}\rangle
\ea
The correlator admits two OPE decompositions which we write
\ba
\label{eq:crossing-eq}
\mathcal G(z) = \frac{1/g}{z^{2\Delta_{\mu}} }\left(1+\sum_{\mathcal{O}} (\lambda^{\mu\mu}_{\mathcal{O}})^2 G_{\Delta_{\mathcal{O}}}(z)\right) = \frac{1/\tilde g}{(1-z)^{2\Delta_\mu}}\left(1+\sum_{\mathcal{\tilde O}} (\lambda^{\mu\mu}_{\mathcal{\tilde O}})^2 G_{\Delta_{\mathcal{\tilde O}}}(1-z)\right).
\ea
where $\Delta_\mu$ is the scaling dimension of both $\mu$ and $\tilde \mu$, and $\cO, \tilde \cO$ stand for operators in the $a$ and $b$ CFTs respectively. We also have introduced the 1d conformal block,
\ba
G_{\Delta}(z)=z^{\Delta}\pFq{2}{1}{\Delta,\Delta}{2\Delta}{z}\:.
\ea
Note that the normalisation of the correlator is slightly different from the one discussed in previous sections so as to make the equation more symmetric between the $a$ and $b$ theories.

Below we will make use of the equality between the two OPE decompositions of the correlator to derive constraints on the CFT data there in.

\subsection{Perturbative bootstrap}
As a first application we consider the case where the theories $a$,$b$ are perturbatively close and related by a short RG flow. Thus we assume that theory $a$ contains a weakly relevant operator $\cO$ with dimension $1-\epsilon$ and conversely that theory $b$ has a weakly irrelevant operator $\tilde \cO$ with dimension $1+\epsilon$. Note that this defines the meaning of $\epsilon$ in the present context.

For $\epsilon=0$, the interface between $a$ and $b$ is trivial, i.e. $\mu=\tilde \mu=\mathds 1$. At small $\epsilon$ we assume that the only operators appearing in the interface OPEs are the identity and $\cO$ (or $\tilde \cO$), so that the crossing equation looks like
\ba
1+(\lambda^{\mu \mu}_{\cO})^2 G_{1-\epsilon}(z)+\ldots=\frac{g}{\tilde g}\left(1+(\lambda^{\mu\mu}_{\tilde \cO})^2 G_{1+\epsilon}\right)+\ldots.
\ea
We can use
\ba
G_{1}(z)=-\log(1-z)\,, \quad \partial_{\Delta} G_{1}(z)=-2\, \text{Li}_2(z)-\log (z) \log (1-z)-2 \log (1-z)\:,
\ea
to find the two parameter solution:
\ba
\Delta_\mu=\Delta_\mu^{(2)}\epsilon^2+\Delta_\mu^{(3)}\epsilon^3 \quad \Rightarrow 
\begin{array}{rl}
(\lambda^{\mu \mu}_{\cO})^2 &=2 \Delta_\mu^{(2)} \epsilon ^2+2 (\Delta_\mu^{(3)}+2
   \Delta_\mu^{(2)}) \epsilon ^3+O\left(\epsilon ^4\right),\vspace{0.5cm}\\
(\lambda^{\mu\mu}_{\tilde \cO})^2 &=2 \Delta_\mu^{(2)} \epsilon ^2+2 (\Delta_\mu^{(3)}-2
   \Delta_\mu^{(2)}) \epsilon ^3+O\left(\epsilon ^4\right),\vspace{0.5cm}\\
   g/\tilde g&=1+\frac{2}{3} \pi ^2
   \Delta_\mu^{(2)} \epsilon ^3+O\left(\epsilon ^4\right).
\end{array}
\ea
This is consistent with the results found in the previous section for $\Delta_\mu$ and the $g$-ratios for the \lrni\ interfaces; and for the $\tt N/D$ interface where $\Delta_\mu^{(3)}=0$. The resulting correlator also agrees with the perturbative computation in appendix \ref{app:perturbative-double-inerface}.

Starting from $O(\epsilon^4)$ an infinite number of operators will contribute to the bootstrap equation and crossing is not sufficiently powerful to fix their individual CFT data. But it is powerful enough to constrain them, and it is to this we turn next.

\subsection{Numerical setup}

We start from the crossing equation~\eqref{eq:crossing-eq} and act with linear functionals on both sides. Let us denote functionals which act on the symmetric and antisymmetric (in $z$) pieces of the crossing equation as $\omega^{+}$ and $\omega^{-}$. A good basis of such functionals 
are the set $\alpha_n^\pm,\beta_n^\pm$ introduced in~\cite{Mazac:2016qev,Mazac:2018mdx,Mazac:2018ycv,Paulos:2020zxx, Kaviraj:2021cvq}. These act on the symmetric (+) and antisymmetric (-) pieces of the crossing equation with respect to $z\leftrightarrow 1-z$. Consider an arbitrary linear combination of such functionals, denoted $\Lambda$. Then a generic crossing constraint can be written
\ba
\Lambda(0)+\sum_{\cO} a_{\cO} \Lambda(\Delta_{\cO})+\frac{g}{\tilde g}\tilde \Lambda(0)+\frac{g}{\tilde g}\sum_{\tilde \cO} a_{\tilde \cO} \tilde \Lambda(\tilde \Delta_{\cO})=0\:,
\ea
where we have set e.g. $a_{\mathcal{O}} = \lambda_{\mu\mu\mathcal{O}}^2$. 
The tilded functional action $\tilde \Lambda$ is defined by setting
\ba
\tilde \omega^{\pm}(\Delta)=\mp \omega^{\pm}(\Delta)\:.
\ea

Below we will be interested in bounding the $g$-factor ratio as a function of $\Delta_\mu$ and other parameters. To do this consider a general linear functional satisfying
\begin{itemize}
\item $\tilde \Lambda(0)=\pm 1$,
\item $\Lambda(\Delta)\geq 0$ for $\Delta \in S$,
\item $\tilde \Lambda(\Delta)\geq 0$ for $\Delta \in \tilde S$.
\end{itemize}
For functionals satisfying these conditions it easily follows from crossing that for CFTs whose spectra lie in $S$ and $\tilde S$ we necessarily have
\ba
\mp\frac{g}{\tilde g}\geq \Lambda(0)\:.
\ea
Thus depending on the sign choice the ratio is lower or upper bounded by $\pm \Lambda(0)$, which we can then ask for to be maximized/minimized.

\subsection{Results}
\subsubsection{The Neumann/Dirichlet interface}
\begin{figure}[H]
    \centering
    \includegraphics[height=7cm]{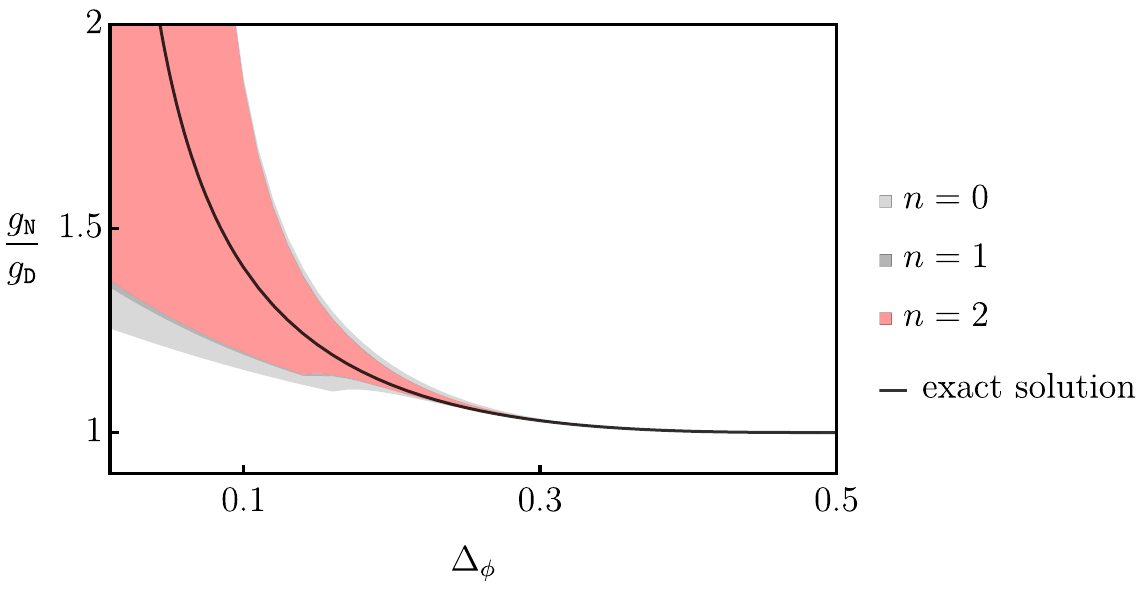}
    \caption{Numerical bounds on $g_{\tt N}/g_{\tt D}$ in function of $\Delta_\phi=\Delta$, with up to $n=2$ functionals, and compared with the exact solution.}
    \label{fig:N/D-num-g}
\end{figure}

As a first computation, we consider a setup corresponding to the Neumann/Dirichlet interface. We use the known interface dimension $\Delta_{\mu}=\frac1{16}(1-2\Delta_\phi)^2$ and define the sets 
\begin{equation}
    S=\{2\Delta_\phi\}\cup[4\Delta_\phi,\infty)\:, \qquad \tilde{S}=\{2-2\Delta_\phi\}\cup[4-4\Delta_\phi,\infty)\:.
\end{equation}
Numerical optimisation in the space of functionals $\alpha_{k\leq n},\beta_{k\leq n}$ leads to the bounds on $g$-factor ratio shown in figure~\ref{fig:N/D-num-g}.
The bounds are sharp in a significant range of $\Delta$ close to $\frac12$. This is consistent with our exact solution for the correlator determined in perturbation theory, which uniquely fixes the ratio to $O\big[(1-2\Delta)^3\big]$. Eventually the bounds widen, but we see they remain compatible with the exact non-perturbative result for the $\tt N/D$ interface \reef{gnonpert}. The upper bound diverges at a finite $\Df$ sufficiently small; this is related to the fact that in this limit the Neumann operator $\phi^2$ has scaling dimensions which approaches zero. Some kind of divergence is indeed necessary, since the exact $g$-factor ratio satisfies
\ba
\frac{g_{\tt N}}{g_{\tt D}} \underset{\Delta\to 0}\sim \frac 1{\sqrt{2\pi \Delta}}\:. \label{eq:gdiv}
\ea

\subsubsection{The Neumann/Long Range Ising interface}
We now consider the Neumann/{\tt LRI} case, the tilded theory being the latter. We set 
\ba
S=\{2\Df\} \cup [4\Df,\infty)\,, \qquad \tilde S=\{\Delta^{\tt LRI}_{\phi^2}\}\cup [\Delta^{\tt LRI}_{\phi^4},\infty)\:.
\ea
The dimension of the $\tt LRI$ operators are taken to be the Pad\'e resummations given in equations \reef{eq:Pade-phi2-phi4}. 

The other important input is $\Delta_\mu$. There are two possibilities. We can use the $O(\epsilon^4)$ result \reef{eq:dmu-NLRI} as an input, so that we have upper and lower bounds as a function of $\Delta_\phi$. The~disadvantage for this is of course that the perturbative result breaks down for sufficiently small $\Df$ so the bounds may not be accurate. The other possibility is to remain agnostic and derive bounds as a function of both $\Df$ and $\Delta_\mu$. In this case the bounds are valid but we don't know where the theory should lie.

\begin{figure}[H]
    \centering
    \includegraphics[height=7cm]{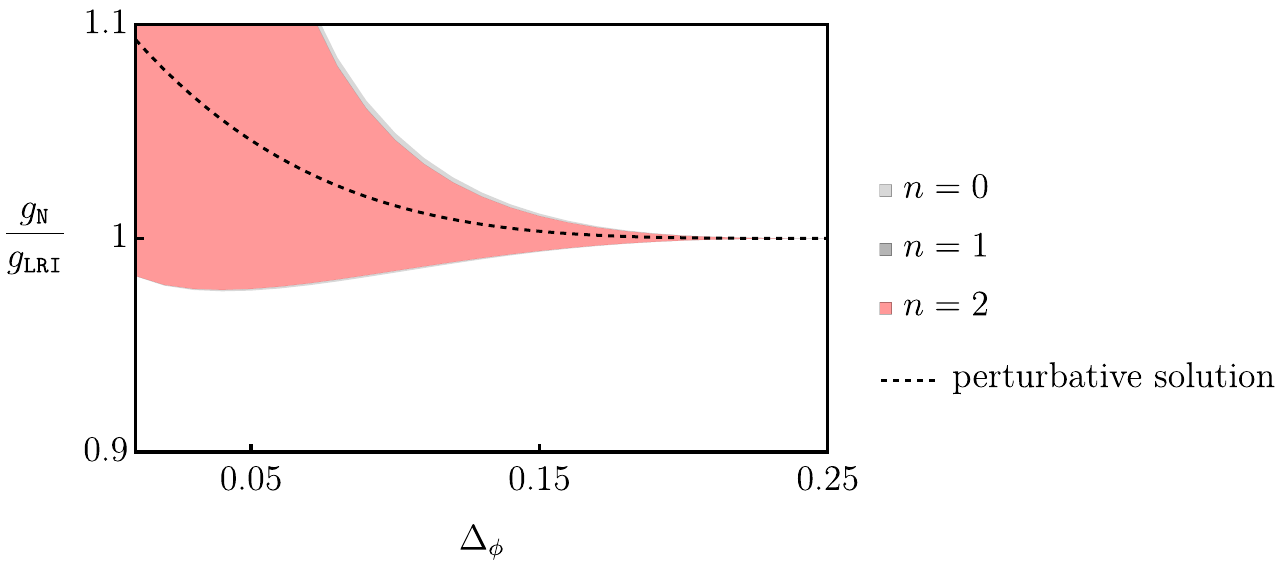}
    \caption{Numerical bounds on $g_{\tt N}/g_{\tt LRI}$ with up to $n=2$ functionals, compared with the perturbative solution to $O(\epsilon^4)$. }
    \label{fig:N/LRI-num-g}
\end{figure}

We present the results for the first approach in figure \ref{fig:N/LRI-num-g}.

The bounds are once again sharp in the vicinity of $\Delta=\frac14$ (or $\epsilon=0$), once again in agreement with our general perturbative solution to order $\epsilon^3$. At higher order the Neumann and {\tt LRI} $\phi^2$ operators necessarily appear in the exact solution, but we don't know with which coefficients. These additional low energy degrees of freedom explain why the bounds are not as tight as in the $\tt N/D$ case. 
The lower bound is evidently not optimal, since the ratio is necessarily larger than one\footnote{This follows from positivity of the righthand side of \reef{formulag}}. The upper bound once again diverges at finite $\Df$ sufficiently small, for the same reason as before. 

\begin{figure}[H]
    \centering
    \includegraphics[height=7cm]{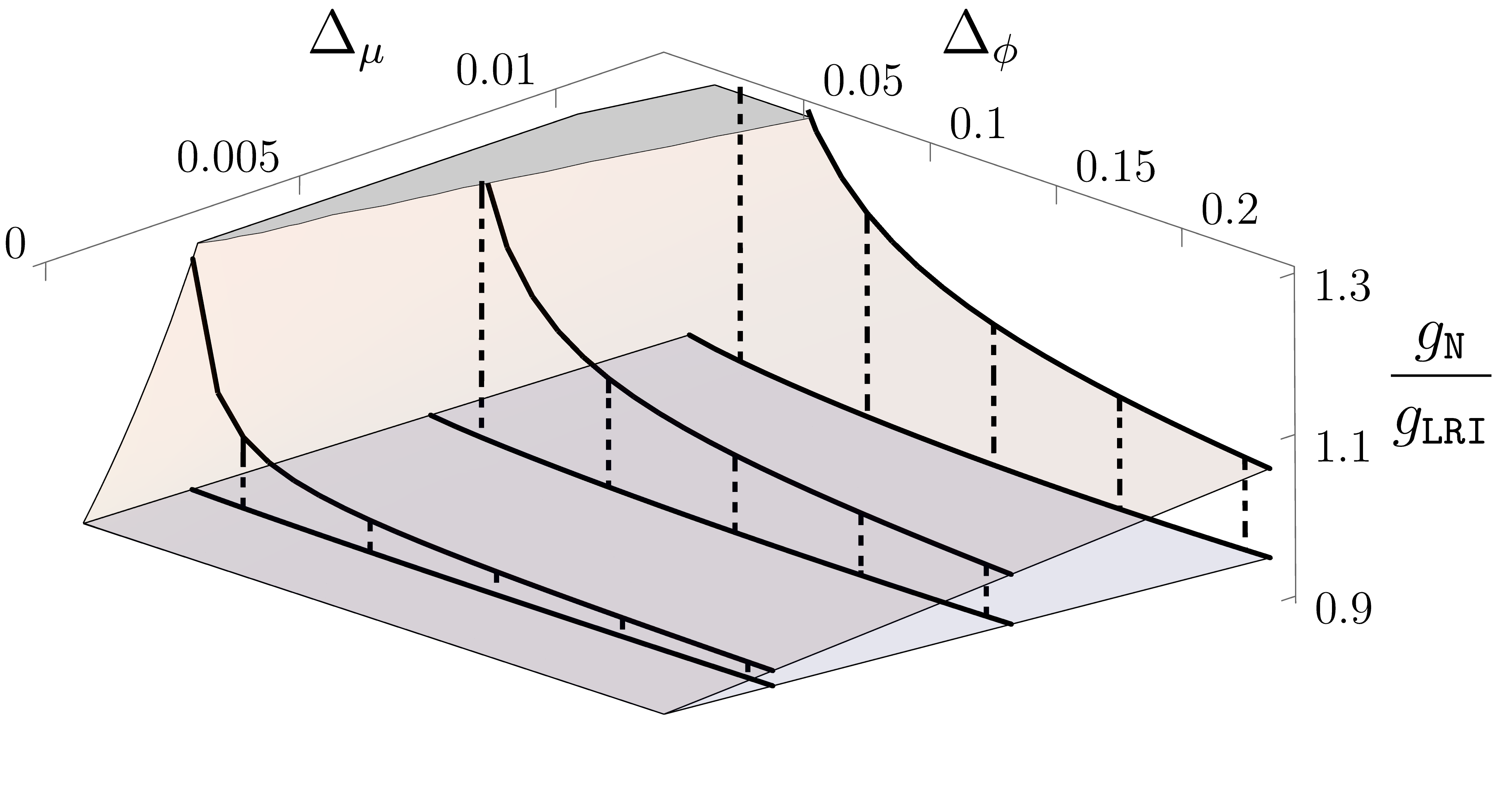}
    \caption{Numerical bounds on $g_{\tt N}/g_\text{LRI}$ with $n=2$ functionals, for general $\Delta_\mu$ and $\Delta_\phi$. The solid lines serve merely to guide the eye.}
    \label{fig:N/LRI-num-g-mu/phi}
\end{figure}

As mentioned above, our bounds are not really valid for small $\Df$ (although this is somewhat irrelevant due to this divergence). We show general bounds depending on $\Df$ and $\Delta_\mu$ in figure \ref{fig:N/LRI-num-g-mu/phi}.

\subsubsection{The Dirichlet/Long Range Ising interface}

\begin{figure}[t]
    \centering
    \includegraphics[height=6.5cm]{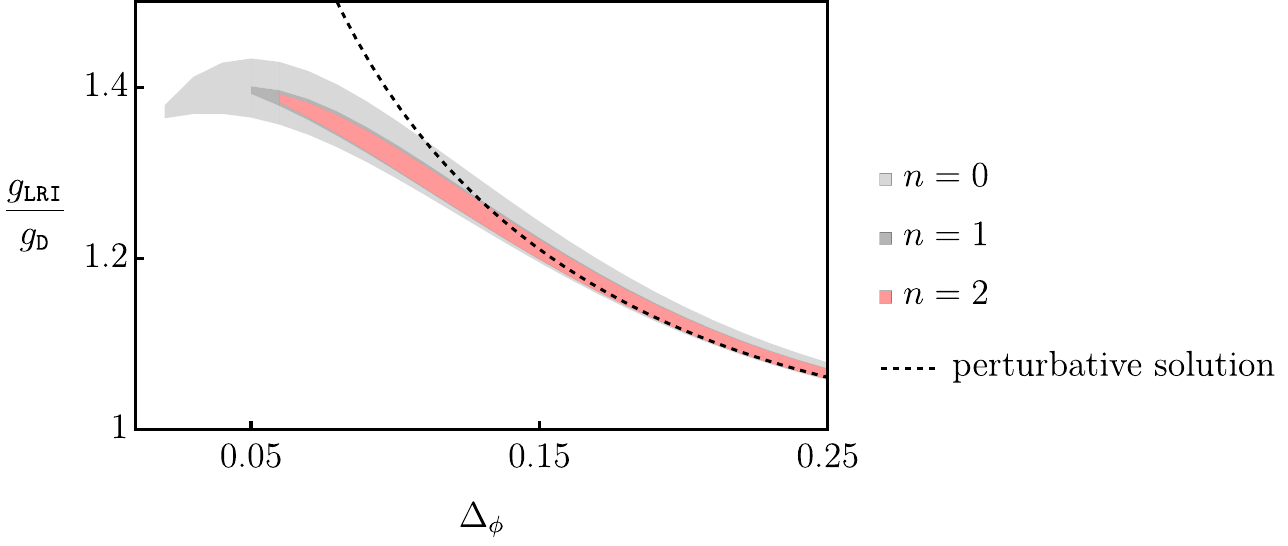}
    \caption{Numerical bounds on $g_{\tt LRI}/g_{\tt D}$ with up to $n=2$ functionals. The bounds assume the perturbative relation $\Delta_\mu^{\tt D/LRI}(\Df)$ and Pad\'e approximations for the dimensions of $\phi^2$ and $\phi^4$ in the LRI channel. The perturbative $g$-factor ratio is valid to $O(\epsilon^4)$ and determined as explained in the main text. The lower bound is approximately saturated by the $\tt N/D$ interface for $\Delta_\phi=\frac1{4}$.}
    \label{fig:D/LRI-num-g}
\end{figure}

\begin{figure}[H]
    \centering
    \includegraphics[height=8cm]{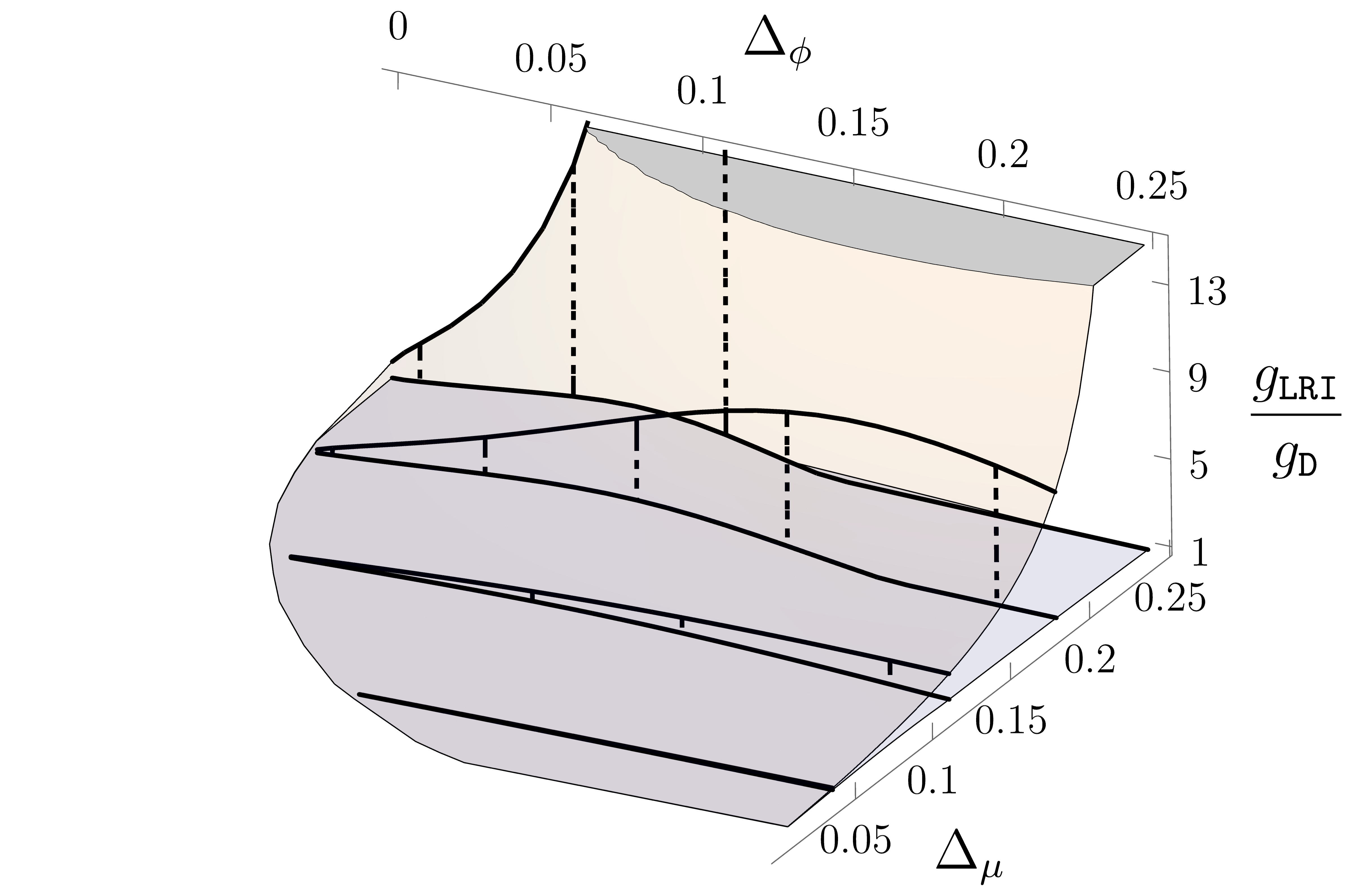}
    \caption{Numerical bounds on $g_\text{LRI}/g_{\tt D}$ with up to $n=2$ functionals, and $\Delta_\mu$/$\Delta_\phi$ decoupled.}
    \label{fig:D/LRI-num-g-mu/phi}
\end{figure}

As our final example, we consider the $\tt D/LRI$ interface. The strategy is the same as in the previous case: we input the known Dirichlet spectrum in one channel and the approximate results in the LRI sector,
\ba
S=\{2-2\Df)\} \cup [4-4\Df,\infty)\,, \qquad \tilde S=\{\Delta^{\tt LRI}_{\phi^2}\}\cup [\Delta^{\tt LRI}_{\phi^4},\infty)\:.
\ea
We again have two possibilities depending on whether we use the perturbative value for $\Delta_\mu$ determined in \reef{eq:dmu-DLRI}. The results are shown in figures \ref{fig:D/LRI-num-g} and \ref{fig:D/LRI-num-g-mu/phi}.
Let us discuss the first plot. It includes both bounds and a perturbative result, obtained by expressing
\ba
\frac{g_{\tt LRI}}{g_{\tt D}}=\frac{g_{\tt N}/g_{\tt D}}{g_{\tt N}/g_{\tt LRI}}\:, \label{ratio}
\ea
and using the exact result \reef{gnonpert} for the numerator on the righthand side and the approximate result \reef{eq:gnglripert} for the denominator.

The bounds can certainly be trusted around $\Delta=\frac14$, but here they are not sharp: this is because at $\Delta=\frac14$ there is already a non-trivial $\tt D/N$ interface, and our perturbative computation of the correlator is certainly invalid. Nevertheless the lower bound does seem to closely follow the perturbative result for the $g$-ratio.

Strikingly, the bounds remain relatively tight and even form a kind of peninsula. This does exclude the small $\Df$ region where the interface certainly exists, but we must remember that we are inputting an approximate expression for $\Delta_\mu$ to derive this bound, which breaks down as we go away from $\Df=\frac 14$. 

Thus we are led to compute bounds without necessarily assuming the value of $\Delta_\mu$, leading to the 3D plot \ref{fig:D/LRI-num-g-mu/phi}.

\begin{figure}[t]
    \centering
    \includegraphics[height=9cm]{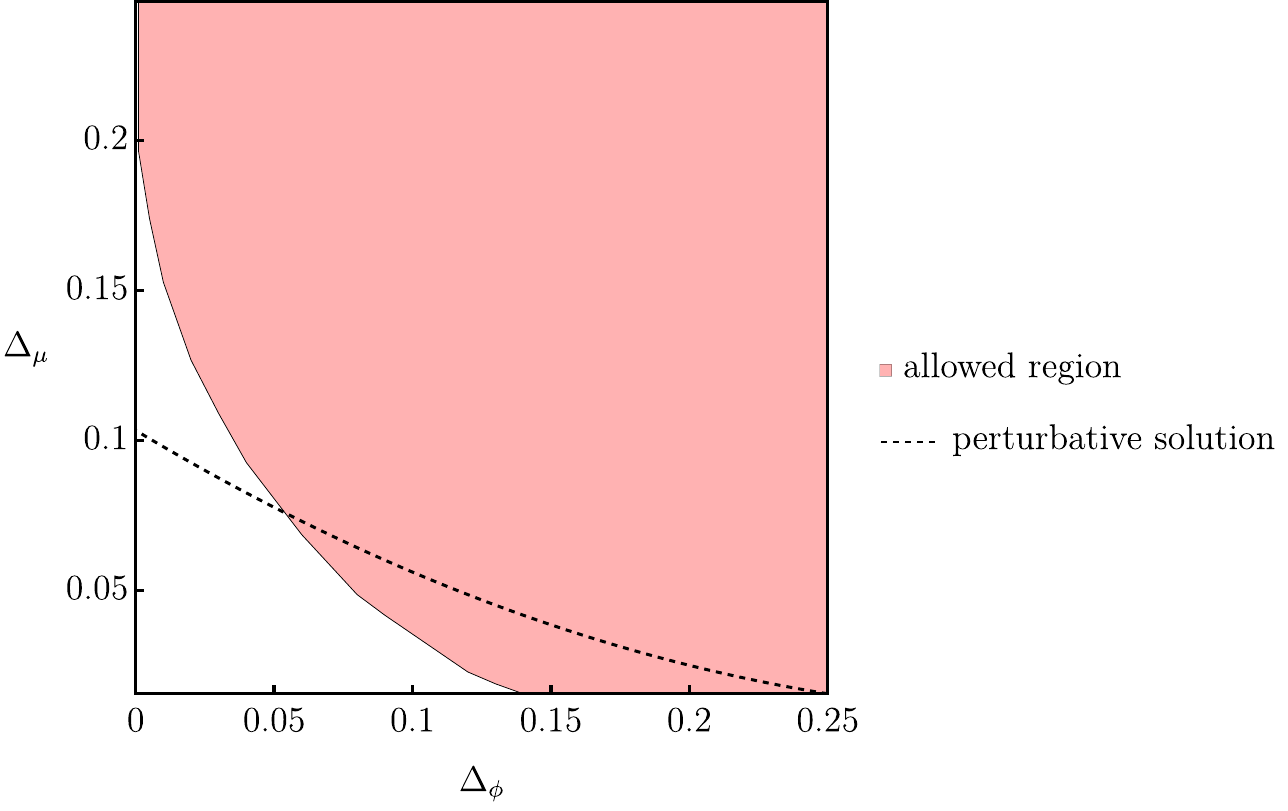}
    \caption{Numerical lower bound on $\Delta_\mu^{\tt D/LRI}(\Delta_\phi)$. The perturbative result $\Delta_\mu^{\tt D/LRI}$ crosses this bound.}
    \label{fig:D/LRI-num-mu/phi}
\end{figure}

\begin{figure}[t]
    \centering
    \includegraphics[height=8cm]{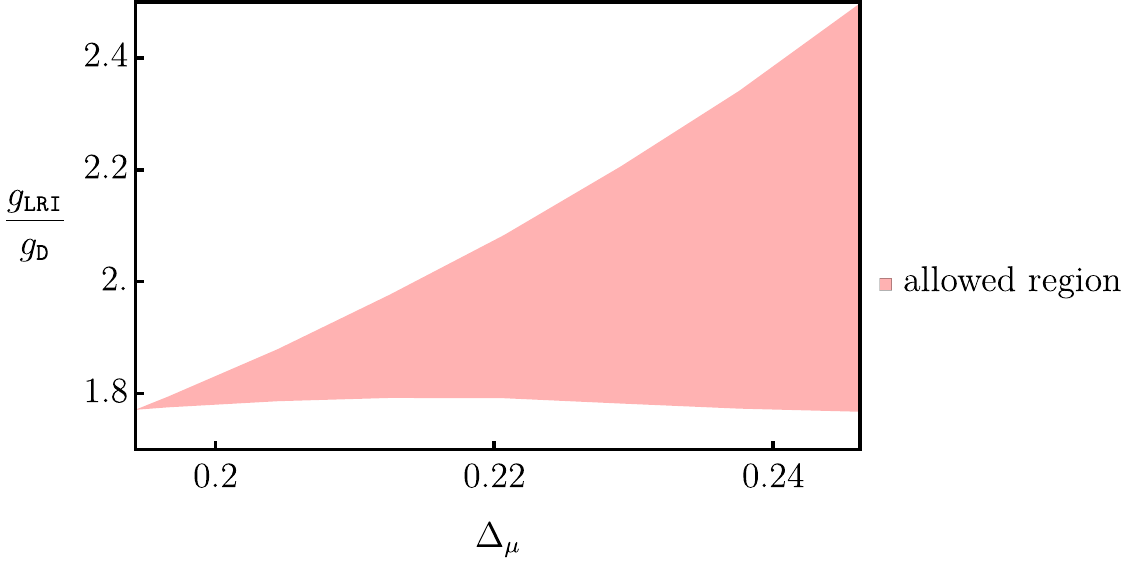}
    \caption{Numerical bounds on $g_{\tt LRI}/g_{\tt D}$ at $\Delta_\phi = 10^{-3}$ in function of $\Delta_\mu^{\tt D/LRI}$. The tip is located at approximately (0.194;1.77)}
    \label{fig:D/LRI-num-mu/g}
\end{figure}

The plot shows that the bounds allow for $\Df=0$ when $\Delta_\mu$ is around 0.2.
One of the consequences of this plot is that there is a lower bound for $\Delta_\mu$ as a function of $\Delta_\phi$, which we show in figure~\ref{fig:D/LRI-num-mu/phi}. The other consequence is that in all likeliness it implies that $g_{\tt N}/g_{\tt LRI}$ is finite for any $\Df$ (and bounded), at least as long as the actual value of $\Delta_\mu$ does not grow too large. We show these bounds for $\Df=0$ in figure~\ref{fig:D/LRI-num-mu/g}. Note that using the trivial rewriting \reef{ratio} together with the divergence \reef{eq:gdiv} of $g_{\tt N}/g_{\tt D}$ for small $\Df$ we get:
\ba
\frac{g_{\tt LRI}}{g_{\tt D}} \quad \mbox{is finite} \qquad \Longrightarrow \quad
\frac{g_{\tt N}}{g_{\tt LRI}}\underset{\Df\to 0} \sim O\big(\Df^{-\frac 12}\big)\:.
\ea
This divergence is nicely compatible with the divergent upper bound we found in that case.

\section{Discussion}
\label{sec:conclusion}
In this paper we have examined interfaces between one-dimensional conformal theories arising from quantum field theories in AdS$_2$. We have stuck to the simplest setting where this QFT is a single, free, scalar field. 

The simplest interfaces correspond to transitions between Neumann and Dirichlet boundary conditions. In this setting we have written down a simple determinant formula for multi-interface correlators, cf. \ref{intcorr}. Similar determinant representations have a long history in the study of correlation functions in integrable models, most notably in the two-dimensional Ising model, where spin correlation functions can be expressed in terms of Fredholm or Toeplitz determinants and related to Painlevé transcendents and isomonodromic tau functions. In upcoming work we will show that here as well it is actually possible to evaluate these determinants explicitly, leading to exact formulae for multi-interface correlators in terms of tau-functions for certain isomonodromic systems.

We have begun a numerical bootstrap exploration of interface correlators but it is clear that it is possible to go much further. In particular it would be important to consider mixed correlator setups exploiting full knowledge of the theory in the Neumann sector.

There are of course many possible generalisations of our setup:  introducing extra fields, fermions, bulk interactions... A particularly nice setup is to consider a $O(N)$ extension. At~the boundary this may be spontaneously and partially broken, and we can then introduce interfaces connecting different breaking patterns. We look forward to examining some of these extensions in the near future.

\subsection*{Acknowledgements}
It is a pleasure to acknowledge discussions with Ant\'onio Antunes, Leonardo Cardinale
 and Philine van Vliet. This work was co-funded by the European Union (ERC, FUNBOOTS, project number 101043588). Views and opinions expressed are however those of the author(s) only and do not necessarily reflect those of the European Union or the European Research Council. Neither the European Union nor the granting authority can be held responsible for them.

\newpage
\appendix
\section{Conformal correlators and crossing}
\label{gff}
In this appendix we collect useful formulae for 1d conformal correlators.
In our conventions, four-point defect correlators are generically written as 
\ba
    &\langle \cO_1(x_1) \cO_2 (x_2) \cO_3 (x_3) \cO_4(x_4)\rangle=\mathcal K_{1234} (x) {\mathcal{G}}_{1234}(z)\:, \\
    &\mathcal K_{1234} (x) = \frac{1}{{x}_{13}^{\Delta_1 + \Delta_3} {x}_{24}^{\Delta_2 + \Delta_4}} \left(\frac{x_{34}}{x_{14}}\right)^{\Delta_{13}} \left(\frac{{x}_{14}}{{x}_{12}} \right)^{\Delta_{24}}\:, \quad x_{ij}=x_i-x_j,\quad \Delta_{ij} = \Delta_i - \Delta_j\:, \label{eq:corr}
\ea
where
\ba
    z=\frac{x_{12} x_{34}}{x_{13} x_{24}}\:. \label{eq:zzbardef}
\ea
In our conventions the crossing equations take the form:
\ba
    {\mathcal{G}}_{1234}(z)&=\left(\frac{z}{1-z}\right)^{\Delta_{24}}{\mathcal G}_{4123}(1-z) \:,\\
    \Leftrightarrow 
    \sum_{\cO} {\lambda}_{12\cO} \lambda_{43\cO} G^{12,34}_{\Delta_\cO}(z,\bar z)&=\left(\frac{z}{1-z}\right)^{\Delta_{24}}\,\sum_{\cO} \lambda_{41\cO} \lambda_{32\cO} G^{41,23}_{\Delta_\cO}(1-z)\:\,,
\ea
where in the second line we used the OPE to express the correlator in terms of conformal blocks, which depend on the conformal dimension ${\Delta}$:
\begin{equation}
    G_{\Delta}^{12,34} = z^{\Delta-\Delta_1-\Delta_4} \,   _2 F_1 \Big(\Delta - \Delta_{12}, \Delta + \Delta_{34}, 2 \Delta; z \Big)\:, \quad \Delta_{ij} = \Delta_i - \Delta_j\:.
\end{equation}
As an example let us consider a GFF of dimension $\Delta$. The four point function can be written as
\ba
\langle \phi|\phi(1)\phi(z)|\phi\rangle&=z^{-2 \Delta}+(1-z)^{-2\Delta}+1\\
&=z^{-2 \Delta}+\sum_{n=0}^\infty \left(\lambda^{\phi \phi}_{(\phi^2)_n}\right)^2  G_{2\Delta+2n}(z)\:,
\ea
where
\ba
\left(\lambda^{\phi \phi}_{(\phi^2)_n}\right)^2=
   \frac{\sqrt{\pi } 2^{-4 \Delta -4 n+3} \Gamma (2
   (n+\Delta )) \Gamma (2 n+4 \Delta -1)}{\Gamma (2
   \Delta )^2 \Gamma (2 n+1) \Gamma \left(2 n+2
   \Delta -\frac{1}{2}\right)}\:.
\ea

\section{Perturbative computations for $\tt N/D$ interface}
\label{app:perturbative}

\subsection{Setup and renormalization}
\label{app:perturbative-setup}

In this appendix, we consider the Neumann/Dirichlet interface perturbatively. We start off with GFF with dimension $\Delta<1/2$, i.e. with Neumann b.c.'s. The operator $:\phi^2:$ is then relevant. Adding it triggers a flow which can be tuned to end at the Dirichlet boundary condition. We can do computations perturbatively by setting
\ba
\epsilon:=1-2\Delta\:,
\ea
and by deforming the action with
\ba
S = S_0 + \frac{g_0}2 \int \ud x \,:\phi^2(x):\:,
\ea
with the integral localized on half of the boundary. We renormalize the coupling $g_0=Z_g g$ and the field $\phi=Z_\phi\tilde{\phi}$. We consider $\phi$ to be canonically normalized, with free propagator
\ba
\begin{tikzpicture}[baseline,valign]
\draw[thick] (-0.1, 0) -- (0.9, 0);
\draw (0.7,0) arc (20:160:0.33);
\end{tikzpicture} 
\;\, \equiv \;
\langle \phi(x)\phi(0) \rangle_{g_0=0}
\, = \;
(x)^{-2\Delta}=|x|^{\epsilon-1} \, ,
\label{eq:free-prop}
\ea 
and the vertex is extracted from the action to be
\ba
\begin{tikzpicture}[baseline,valign]
\draw[thick] (-0.1, 0) -- (0.7, 0);
\draw(0.3, 0) -- (0.15, 0.6);
\draw(0.3, 0) -- (0.45, 0.6);
\node at (0.3, 0) [dcirc] {};
\end{tikzpicture}
\;\; \equiv \;
- \frac{g_0}2 \int\mathrm{d} x \, .  
\ea

For the coupling renormalization, we need to determine the counter-terms to the correction $-g_0 \int \ud x \,:\phi^2(x):\,$. They come from the diagrams
\ba
\begin{tikzpicture}[baseline,valign,scale=1.5]
\draw[thick] (-0.1, 0) -- (1.1, 0);
\draw[thick, dashed] (1.1, 0) -- (1.6, 0);
\draw[thick] (1.6, 0) -- (2.3, 0);

\draw (0.6,0) arc (20:160:0.26);
\draw (1.1,0) arc (20:160:0.26);
\draw (1.1,0) arc (20:160:0.26);
\draw[dashed] (1.6,0) arc (20:160:0.26);
\draw (2.1,0) arc (20:160:0.26);
\draw (2.1,0) arc (-55:-125:1.74);
\node at (0.1, 0) [dcirc] {};
\node at (0.6, 0) [dcirc] {};
\node at (1.1, 0) [dcirc] {};
\node at (1.6, 0) [dcirc] {};
\node at (2.1, 0) [dcirc] {};
\end{tikzpicture}
\;\; = \;
\frac{(-g_0)^n}{2^n n!} \int\cdots\int\ud x_1\cdots\ud x_n \,\phi^2(x_1)\cdots\phi^2(x_n)\:.
\ea

At 1-loop, the integral is simply
\ba
\iint \ud x_1\ud x_2\, \phi^2(x_1) \phi^2(x_2)\supset 4 \int \ud x_1\, \phi^2(x_1)\int\ud x_2\,\frac{1}{x_{12}^{1-\epsilon}}=\frac{8}{\epsilon} \int \ud x_1 \,\phi^2(x_1)+\mathcal{O}(1).
\ea

All the other higher-loop integrals follow directly. After taking care of the ordering and the normalization, we get counter-terms
\ba
Z_g = M^{\epsilon}\sum_{n=0}^\infty\left(\frac{2g}\epsilon\right)^n = \frac{M^{\epsilon}}{1-\frac{2g}\epsilon}\:,
\ea
with $M$ the renormalization scale. The dependence of the renormalized coupling on this scale gives the beta function
\ba
\beta(g)=-\epsilon g + 2g^2\:.
\ea

Its nontrivial fixed point is at $g^\star=\frac 12\epsilon$. For the field renormalization, the $n$-loop counter-term comes from the diagram
\ba
\langle\phi(x)\phi(0)\rangle^{(n)} &= \begin{tikzpicture}[baseline,valign,,scale=1.5]
\draw[thick] (-0.1, 0) -- (1.1, 0);
\draw[thick, dashed] (1.1, 0) -- (1.6, 0);
\draw[thick] (1.6, 0) -- (2.3, 0);
\draw (0.6,0) arc (20:160:0.26);
\draw (1.1,0) arc (20:160:0.26);
\draw (1.1,0) arc (20:160:0.26);
\draw[dashed] (1.6,0) arc (20:160:0.26);
\draw (2.1,0) arc (20:160:0.26);
\node at (0.6, 0) [dcirc] {};
\node at (1.1, 0) [dcirc] {};
\node at (1.6, 0) [dcirc] {};
\end{tikzpicture} \\
&= \;
\frac{(-g_0)^n}{2^n n!} \langle\phi(x)\left(\int\cdots\int\ud x_1\cdots\ud x_n \,\phi^2(x_1)\cdots\phi^2(x_n)\right)\phi(0)\rangle\:.
\ea

These can be computed recursively the integral by using the standard chain integral
\ba
\int \frac{\ud x}{(x-x_1)^a(x-x_2)^b} = \frac{\sqrt{\pi}}{(x_{12})^{a+b-\frac12}}\frac{\Gamma\big(\frac12-a\big)\Gamma\big(\frac12-b\big)\Gamma\big(a+b-\frac12\big)}{\Gamma(a)\Gamma(b)\Gamma(1-a-b)}\:.
\ea

The product of ratios of gamma functions is telescopic, simplifying greatly to
\ba 
\langle \phi(x) \phi(0)\rangle^{(n)} = \frac{(-g_0)^n}{x^{1-(n+1)\epsilon}}\pi^{n/2}\left(\frac{\Gamma\big(\tfrac\epsilon2\big)}{\Gamma\big(\tfrac{1-\epsilon}2\big)}\right)^{n+1}\frac{\Gamma\big(\tfrac{1-(n+1)\epsilon}2\big)}{\Gamma\big(\tfrac{(n+1)\epsilon}2\big)}\:.
\ea

Cancelling the poles, we find the field renormalization, that can be checked at very high-order to be simply
\ba
Z_\phi&=1-\frac{2g}{\epsilon}\:.
\ea

The conformal dimension at the fixed point $g^\star=\frac12\epsilon$ is then
\ba
\tilde{\Delta}=\Delta+\epsilon=\frac{1+\epsilon}2\:,
\ea
thus yielding the expected relation $\Delta_{\tt D}=1-\Delta_{\tt N}$. One final conformal data that can be computed is the field normalization
\ba
\mathcal{N}_\phi=1+\epsilon^2\frac{\pi^2}{12}+\epsilon^4\frac{\pi^4}{120}+\cdots
\ea

\subsection{Alternative computation in momentum space}

The previous computation in position space is useful to set the basics for more intricate correlators, but going to momentum space proves in this case to be more direct. The Fourier transform of the free two-point correlation function is
\ba
G_0(\omega) = \int \ud x\, e^{i\omega x}|x|^{\epsilon-1} = 2\Gamma(\epsilon)\cos\big(\tfrac12\pi\epsilon\big) |\omega|^{-\epsilon}\:.
\ea
The double-trace perturbation in momentum space gives a geometric series
\ba
G(\omega) = G_0(\omega) - g_0 G_0^2(\omega) + g_0^2G_0^3(\omega) - \cdots = \frac{G_0(\omega)}{1+g_0 G_0(\omega)}\:.
\ea
In the IR limit $g_0G_0(\omega) \gg 1$, it expands as
\ba
G(\omega) = \frac1{g_0} - \frac1{g_0^2 G_0(\omega)} + \cdots
\ea
The first term is simply a contact term. Fourier-transforming back to position space the $|\omega|^\epsilon$ of the leading non-local contribution
\ba
\int \frac{\ud \omega}{2\pi}\, e^{-i\omega x}|\omega|^{\epsilon} = -\tfrac1\pi\Gamma(1+\epsilon)\sin\big(\tfrac12\pi\epsilon\big) |x|^{-\epsilon-1}\:.
\ea
Going to the fixed point $g^\star=\frac12\epsilon$, we get the two-point correlation function
\ba
\langle \tilde \phi(x)\tilde \phi(0)\rangle = \frac2{\epsilon\pi}\tan\left(\frac{\pi\epsilon}2\right) |x|^{-\epsilon-1}\:.
\ea
This reproduces the previous result $\Delta_{\tt D}=1-\Delta_{\tt N}$, and the normalization has an expansion that matches the one found in position space.

\subsection{Interface energy \texorpdfstring{$\Delta_\mu^{\tt N/D}$}{dmu}}
\label{app:perturbative-dmu}

The next observable we can compute perturbatively is the interface energy $\Delta_\mu^{\tt N/D}$. For this we consider an interface from 0 to $T$, and we compute the euclidean partition function on the cylinder
\ba
Z^{\tt N/D}(T)\equiv \langle\mu(0)\mu(T)\rangle=\langle e^{-g_0 \int_0^T \ud x\,\phi^2(x)} \rangle \:.
\ea

The diagrams are exactly the ones considered for the coupling renormalization, except that here the bounds of the integral are fixed, making the computations more intricate. Up to three loop, all the permutations of the insertions are equivalent, and the diagrams all have the general expression
\ba
\begin{tikzpicture}[baseline,valign,scale=2]
\draw[thick] (-0.1, 0) node[below, yshift=-5pt] {0} -- (1.1, 0) ;
\draw[thick] (-0.1, -0.1) -- (-0.1, 0.1);
\draw[thick, dashed] (1.1, 0) -- (1.6, 0);
\draw[thick] (1.6, 0) -- (2.3, 0) node[below, yshift=-5pt] {$T$};
\draw[thick] (2.3, -0.1) -- (2.3, 0.1);
\draw (0.6,0) arc (20:160:0.26);
\draw (1.1,0) arc (20:160:0.26);
\draw (1.1,0) arc (20:160:0.26);
\draw[dashed] (1.6,0) arc (20:160:0.26);
\draw (2.1,0) arc (20:160:0.26);
\draw (2.1,0) arc (-55:-125:1.74);
\node (x1) at (0.1, 0) [dcirc] {};
\node (x2) at (0.6, 0) [dcirc] {};
\node (x3) at (1.1, 0) [dcirc] {};
\node (xnm1) at (1.6, 0) [dcirc] {};
\node (xn) at (2.1, 0) [dcirc] {};
\draw (x1) node [above, yshift=5pt]{\small$x_1$};        
\draw (x2) node [above, yshift=5pt]{\small$x_2$};        
\draw (x3) node [above, yshift=5pt]{\small$x_3$};        
\draw (xnm1) node [above, yshift=5pt]{\small$x_{n-1}$};        
\draw (xn) node [above, yshift=5pt]{\small$x_n$};        
\end{tikzpicture} &= (-g_0)^n\int_0^T\ud x_1\int_0^{x_1}\frac{\ud x_2}{(x_{12})^{{1-\epsilon}}} \cdots \int_0^{x_{n-1}}\frac{\ud x_n}{(x_{n-1,n})^{{1-\epsilon}}}\frac{1}{(x_{1n})^{{1-\epsilon}}}\\
&= -\frac{(-g_0)^n\Gamma(\epsilon)^{n-1}T^{n\epsilon}}{n\epsilon(1-n\epsilon)\Gamma\big((n-1)\epsilon\big)}\:.\label{integrals}
\ea

After taking care of the ordering factors and the normalization, we get
\ba
\langle\mu(0)\mu(T)\rangle=1-g_0^2 \frac{T^{2\epsilon}}{\epsilon(1-2\epsilon)}+ 8g_0^3\frac{\Gamma (\epsilon)^2}{3 \epsilon 
(1-3 \epsilon) \Gamma \left(2\epsilon\right)}T^{3\epsilon }\:.
\ea

The renormalization can be extracted, from which we can then compute the conformal dimension at the fixed point
\ba
Z_\mu &= 1-\frac{g_0^2}{2\epsilon}+\frac{8g_0^3}{3\epsilon^2}+\mathcal{O}(\epsilon^4), \\
\Delta_\mu^{\tt N/D} &= \frac{\epsilon^2}{16}+\mathcal{O}(\epsilon^4)\:.
\ea

We recover the non-perturbative result $\Delta_\mu^{\tt N/D}=\frac 14 (\Delta-\frac 12)^2$. Finally, one can compute the normalization
\ba
\mathcal{N}_\mu=1-\frac{\epsilon^2}{8}-\frac{\pi^2\epsilon^3}{72}+\mathcal{O}(\epsilon^4)\:.
\ea

\subsection{Boundary correlators}
\label{app:perturbative-bcorrelator}

We go on to the boundary correlators
\ba
\langle\mu\phi\phi\mu\rangle &= \begin{tikzpicture}[baseline,valign]
\draw[thick] (-0.05, 0) node[below, yshift=-2pt] {0} -- (1.1, 0) node[below, yshift=-2pt] {$T$};
\draw[thick] (1.1, -0.1) -- (1.1, 0.1);
\draw[thick] (-0.05, -0.1) -- (-0.05, 0.1);   
\draw (0.9,0) arc (20:160:0.4);
\end{tikzpicture} +
\begin{tikzpicture}[baseline,valign]
\draw[thick] (-0.05, 0) node[below, yshift=-2pt] {0} -- (1.85, 0) node[below, yshift=-2pt] {$T$};
\draw[thick] (1.85, -0.1) -- (1.85, 0.1);
\draw[thick] (-0.05, -0.1) -- (-0.05, 0.1);   
\draw (0.9,0) arc (20:160:0.4);
\draw (1.65,0) arc (20:160:0.4);
\node (x1) at (0.9, 0) [dcirc] {};
\end{tikzpicture} +
\begin{tikzpicture}[baseline,valign]
\draw[thick] (-0.05, 0) node[below, yshift=-2pt] {0} -- (2.6, 0) node[below, yshift=-2pt] {$T$};
\draw[thick] (2.6, -0.1) -- (2.6, 0.1);
\draw[thick] (-0.05, -0.1) -- (-0.05, 0.1);   
\draw (0.9,0) arc (20:160:0.4);
\draw (1.65,0) arc (20:160:0.4);
\draw (2.4,0) arc (20:160:0.4);
\node at (0.9, 0) [dcirc] {};
\node at (1.65, 0) [dcirc] {};
\end{tikzpicture} +
\begin{tikzpicture}[baseline,valign]
\draw[thick] (-0.05, 0) node[below, yshift=-2pt] {0} -- (2.1, 0) node[below, yshift=-2pt] {$T$};
\draw[thick] (2.1, -0.1) -- (2.1, 0.1);
\draw[thick] (-0.05, -0.1) -- (-0.05, 0.1);   
\draw (0.9,0) arc (20:160:0.4);
\draw (1.9,0) arc (-20:-160:0.4);
\draw (1.9,0) arc (20:160:0.4);
\node at (1.15, 0) [dcirc] {};
\node at (1.9, 0) [dcirc] {};
\end{tikzpicture} \:, \\
\langle\mu\phi^2\phi^2\mu\rangle &= \begin{tikzpicture}[baseline,valign]
\draw[thick] (-0.05, 0) node[below, yshift=-2pt] {0} -- (1.1, 0) node[below, yshift=-2pt] {$T$};
\draw[thick] (1.1, -0.1) -- (1.1, 0.1);
\draw[thick] (-0.05, -0.1) -- (-0.05, 0.1);   
\draw (0.9,0) arc (20:160:0.4);
\draw (0.9,0) arc (-20:-160:0.4);
\end{tikzpicture} +
\begin{tikzpicture}[baseline,valign]
\draw[thick] (-0.05, 0) node[below, yshift=-2pt] {0} -- (1.85, 0) node[below, yshift=-2pt] {$T$};
\draw[thick] (1.85, -0.1) -- (1.85, 0.1);
\draw[thick] (-0.05, -0.1) -- (-0.05, 0.1);   
\draw (0.9,0) arc (20:160:0.4);
\draw (1.65,0) arc (20:160:0.4);
\draw (1.65,0) arc (-40:-140:0.98);
\node (x1) at (0.9, 0) [dcirc] {};
\end{tikzpicture} +
\begin{tikzpicture}[baseline,valign]
\draw[thick] (-0.05, 0) node[below, yshift=-2pt] {0} -- (2.6, 0) node[below, yshift=-2pt] {$T$};
\draw[thick] (2.6, -0.1) -- (2.6, 0.1);
\draw[thick] (-0.05, -0.1) -- (-0.05, 0.1);   
\draw (0.9,0) arc (20:160:0.4);
\draw (1.65,0) arc (20:160:0.4);
\draw (2.4,0) arc (20:160:0.4);
\draw (2.4,0) arc (-50:-130:1.75);
\node at (0.9, 0) [dcirc] {};
\node at (1.65, 0) [dcirc] {};
\end{tikzpicture} +
\begin{tikzpicture}[baseline,valign]
\draw[thick] (-0.05, 0) node[below, yshift=-2pt] {0} -- (2.1, 0) node[below, yshift=-2pt] {$T$};
\draw[thick] (2.1, -0.1) -- (2.1, 0.1);
\draw[thick] (-0.05, -0.1) -- (-0.05, 0.1);   
\draw (0.9,0) arc (20:160:0.4);
\draw (0.9,0) arc (-20:-160:0.4);
\draw (1.9,0) arc (-20:-160:0.4);
\draw (1.9,0) arc (20:160:0.4);
\node at (1.15, 0) [dcirc] {};
\node at (1.9, 0) [dcirc] {};
\end{tikzpicture} \\
&+ \begin{tikzpicture}[baseline,valign]
\draw[thick] (-0.05, 0) node[below, yshift=-2pt] {0} -- (2.1, 0) node[below, yshift=-2pt] {$T$};
\draw[thick] (2.1, -0.1) -- (2.1, 0.1);
\draw[thick] (-0.05, -0.1) -- (-0.05, 0.1);   
\draw (0.9,0) arc (20:160:0.4);
\draw (0.9,0) arc (-20:-160:0.4);
\draw (1.9,0) arc (-20:-160:0.4);
\draw (1.9,0) arc (20:160:0.4);
\node at (1.15, 0) [dcirc] {};
\node at (0.9, 0) [dcirc] {};
\end{tikzpicture} +
\begin{tikzpicture}[baseline,valign]
\draw[thick] (-0.05, 0) node[below, yshift=-2pt] {0} -- (2.2, 0) node[below, yshift=-2pt] {$T$};
\draw[thick] (2.2, -0.1) -- (2.2, 0.1);
\draw[thick] (-0.05, -0.1) -- (-0.05, 0.1);   
\draw (0.9,0) arc (20:160:0.4);
\draw (1.25,0) arc (-35:-145:0.67);
\draw (2,0) arc (-20:-160:0.4);
\draw (2,0) arc (35:145:0.67);
\node at (1.25, 0) [dcirc] {};
\node at (0.9, 0) [dcirc] {};
\end{tikzpicture} \:.
\ea

We go to the frame $\langle\mu(0)\mu(z)\mathcal{O}(1)\mathcal{O}(\infty)\rangle$. The normalized correlator have the expression
\ba
G_{\mu\mu\phi\phi}(1-z) &= 1 -2 g_0 I_1(\epsilon,z) + 4g_0^2 I_2(\epsilon,z)\:,\\
G_{\mu\mu\phi^2\phi^2}(1-z) &= 1-4 g_0 I_1(\epsilon,z) +g_0^2\big(4I_1(\epsilon,z)^2+8I_2(\epsilon,z)+2I_3(\epsilon,z)\big)\:,
\ea
where we defined the integrals
\ba
I_1(\epsilon,z)&= \int_0^z |1-x|^{\epsilon-1}\,\mathrm{d}x = \frac{1-(1-z)^\epsilon}{\epsilon}\:, \\
I_2(\epsilon,z)&= \iint_0^z |1-x|^{\epsilon-1}|x-y|^{\epsilon-1}\,\mathrm{d}x\mathrm{d}y \\
&= \frac{z^{\epsilon+1}}{\epsilon(\epsilon+1)}\left[{}_2F_1(1-\epsilon,1+\epsilon,2+\epsilon,z)+{}_2F_1(1-\epsilon,1,2+\epsilon,z) \right]\: , \\
I_3(\epsilon,z) &= z\int_0^z |1-x|^{2\epsilon-2}\,\mathrm{d}x = z\frac{(1-z)^{2\epsilon-1}-1}{1-2\epsilon}\:.
\ea

Replacing the results of the integral, we get
\ba
G_{\mu\mu\phi\phi}(1-z) &= 1+\tfrac12\epsilon\log(1-z)+\epsilon^2\left[\tfrac18\log^2(1-z)-\tfrac12\log(1-z)\log z - \tfrac12\text{Li}_2(z) \right]\:, \\
G_{\mu\mu\phi^2\phi^2}(1-z)&= 1+\epsilon\log(1-z)+\epsilon^2\left[\tfrac{z^2}{8(1-z)}+\tfrac12\log^2(1-z)-\log(1-z)\log z - \text{Li}_2(z) \right]\:,
\ea
that agree perfectly to the non-perturbative correlators. One can of course match their block expansion and match the OPE coefficients $\lambda^{\mu\mu}_{(\phi^2)_n}$ and $\lambda^{\mu\mu}_{(\phi^4)_n}$.

\subsection{Double interface and \texorpdfstring{$g$}{g}-ratio}
\label{app:perturbative-double-inerface}

For now, we had only considered a single interface. If we consider a second interface, it amounts to insert a second pair of $\mu$ operators, hence computing a four-point function $\langle\mu\mu\mu\mu\rangle$. Up to three loops, the diagrams involved are
\ba
\langle \mu(\infty) \mu(1) \mu(z)\mu(0)\rangle &~=~ \begin{tikzpicture}[baseline,valign]
\draw[thick] (0, 0) node[below, yshift=-2pt] {0} -- (1, 0) node[below, yshift=-2pt] {$z$};
\draw[thick] (1, -0.1) -- (1, 0.1);
\draw[thick] (0, -0.1) -- (0, 0.1);  
\draw[thick] (0, 0.7) node[above, yshift=2pt] {1} -- (1, 0.7) node[above, yshift=2pt] {$\infty$};
\draw[thick] (1, 0.6) -- (1, 0.8);  
\draw[thick] (0, 0.6) -- (0, 0.8);
\end{tikzpicture} ~+~
\begin{tikzpicture}[baseline,valign]
\draw[thick] (0, 0) node[below, yshift=-2pt] {0} -- (1.2, 0) node[below, yshift=-2pt] {$z$};
\draw[thick] (1.2, -0.1) -- (1.2, 0.1);
\draw[thick] (0, -0.1) -- (0, 0.1);  
\draw[thick] (0, 0.7) node[above, yshift=2pt] {1} -- (1.2, 0.7) node[above, yshift=2pt] {$\infty$};
\draw[thick] (1.2, 0.6) -- (1.2, 0.8);  
\draw[thick] (0, 0.6) -- (0, 0.8);
\draw (1,0) arc (20:160:0.43);
\draw (1,0) arc (-20:-160:0.43);
\node at (1, 0) [dcirc] {};
\node at (0.2, 0) [dcirc] {};
\end{tikzpicture} ~+~
\begin{tikzpicture}[baseline,valign]
\draw[thick] (0, 0) node[below, yshift=-2pt] {0} -- (1.2, 0) node[below, yshift=-2pt] {$z$};
\draw[thick] (1.2, -0.1) -- (1.2, 0.1);
\draw[thick] (0, -0.1) -- (0, 0.1);  
\draw[thick] (0, 0.7) node[above, yshift=2pt] {1} -- (1.2, 0.7) node[above, yshift=2pt] {$\infty$};
\draw[thick] (1.2, 0.6) -- (1.2, 0.8);  
\draw[thick] (0, 0.6) -- (0, 0.8);
\draw (1,0.7) arc (20:160:0.43);
\draw (1,0.7) arc (-20:-160:0.43);
\node at (1, 0.7) [dcirc] {};
\node at (0.2, 0.7) [dcirc] {};
\end{tikzpicture} ~+~
\begin{tikzpicture}[baseline,valign]
\draw[thick] (0, 0) node[below, yshift=-2pt] {0} -- (1.2, 0) node[below, yshift=-2pt] {$z$};
\draw[thick] (1.2, -0.1) -- (1.2, 0.1);
\draw[thick] (0, -0.1) -- (0, 0.1);  
\draw[thick] (0, 0.7) node[above, yshift=2pt] {1} -- (1.2, 0.7) node[above, yshift=2pt] {$\infty$};
\draw[thick] (1.2, 0.6) -- (1.2, 0.8);  
\draw[thick] (0, 0.6) -- (0, 0.8);
\draw (0.6,0.7) arc (130:230:0.45);
\draw (0.6,0.7) arc (-310:-410:0.45);
\node at (0.6, 0) [dcirc] {};
\node at (0.6, 0.7) [dcirc] {};
\end{tikzpicture} \\
&~+~
\begin{tikzpicture}[baseline,valign]
\draw[thick] (0, 0) node[below, yshift=-2pt] {0} -- (1.2, 0) node[below, yshift=-2pt] {$z$};
\draw[thick] (1.2, -0.1) -- (1.2, 0.1);
\draw[thick] (0, -0.1) -- (0, 0.1);  
\draw[thick] (0, 0.7) node[above, yshift=2pt] {1} -- (1.2, 0.7) node[above, yshift=2pt] {$\infty$};
\draw[thick] (1.2, 0.6) -- (1.2, 0.8);  
\draw[thick] (0, 0.6) -- (0, 0.8);
\draw (0.2,0.7) arc (180:240:0.8);
\draw (1,0.7) arc (-360:-420:0.8);
\draw (1,0.7) arc (20:160:0.43);
\node at (0.2, 0.7) [dcirc] {};
\node at (1, 0.7) [dcirc] {};
\node at (0.6, 0) [dcirc] {};
\end{tikzpicture} ~+~
\begin{tikzpicture}[baseline,valign]
\draw[thick] (0, 0) node[below, yshift=-2pt] {0} -- (1.2, 0) node[below, yshift=-2pt] {$z$};
\draw[thick] (1.2, -0.1) -- (1.2, 0.1);
\draw[thick] (0, -0.1) -- (0, 0.1);  
\draw[thick] (0, 0.7) node[above, yshift=2pt] {1} -- (1.2, 0.7) node[above, yshift=2pt] {$\infty$};
\draw[thick] (1.2, 0.6) -- (1.2, 0.8);  
\draw[thick] (0, 0.6) -- (0, 0.8);
\draw (0.2,0) arc (-180:-240:0.8);
\draw (1,0) arc (360:420:0.8);
\draw (1,0) arc (-20:-160:0.43);
\node at (0.2, 0) [dcirc] {};
\node at (1, 0) [dcirc] {};
\node at (0.6, 0.7) [dcirc] {};
\end{tikzpicture}
\ea
The first three are similar as for a single interface. The next two can be computed exactly
\ba
\begin{tikzpicture}[baseline,valign]
\draw[thick] (0, 0) node[below, yshift=-2pt] {0} -- (1.2, 0) node[below, yshift=-2pt] {$z$};
\draw[thick] (1.2, -0.1) -- (1.2, 0.1);
\draw[thick] (0, -0.1) -- (0, 0.1);  
\draw[thick] (0, 0.7) node[above, yshift=2pt] {1} -- (1.2, 0.7) node[above, yshift=2pt] {$\infty$};
\draw[thick] (1.2, 0.6) -- (1.2, 0.8);  
\draw[thick] (0, 0.6) -- (0, 0.8);
\draw (0.6,0.7) arc (130:230:0.45);
\draw (0.6,0.7) arc (-310:-410:0.45);
\node at (0.6, 0) [dcirc] {};
\node at (0.6, 0.7) [dcirc] {};
\end{tikzpicture} &~=~ 2g_0^2 \int_{1}^\infty \ud x \int_0^z \ud x' \frac{1}{(x-x')^{2-2\epsilon}} = \frac{g_0^2}{\epsilon(1-2\epsilon)}\big(1-(1-z)^{2\epsilon}\big)\:,  \\
\begin{tikzpicture}[baseline,valign]
\draw[thick] (0, 0) node[below, yshift=-2pt] {0} -- (1.2, 0) node[below, yshift=-2pt] {$z$};
\draw[thick] (1.2, -0.1) -- (1.2, 0.1);
\draw[thick] (0, -0.1) -- (0, 0.1);  
\draw[thick] (0, 0.7) node[above, yshift=2pt] {1} -- (1.2, 0.7) node[above, yshift=2pt] {$\infty$};
\draw[thick] (1.2, 0.6) -- (1.2, 0.8);  
\draw[thick] (0, 0.6) -- (0, 0.8);
\draw (0.2,0.7) arc (180:240:0.8);
\draw (1,0.7) arc (-360:-420:0.8);
\draw (1,0.7) arc (20:160:0.43);
\node at (0.2, 0.7) [dcirc] {};
\node at (1, 0.7) [dcirc] {};
\node at (0.6, 0) [dcirc] {};
\end{tikzpicture} &~=~ -8g_0^3 \int_1^\infty\int_1^{x_1}\int_0^z \frac{\ud x_1\,\ud x_2\,\ud x_3}{(x_{12}x_{13}x_{23})^{1-\epsilon}} = -16g_0^3\,\frac{\Gamma(\epsilon)\Gamma(-2\epsilon)}{3\epsilon(1-3\epsilon)\Gamma(-\epsilon)}\left(1-(1-z)^{3\epsilon}\right) \:.
\ea

For the last one, we only pick up the pole and the finite part by expanding the integrand in $\epsilon$ and then integrating
\ba
\begin{tikzpicture}[baseline,valign]
\draw[thick] (0, 0) node[below, yshift=-2pt] {0} -- (1.2, 0) node[below, yshift=-2pt] {$z$};
\draw[thick] (1.2, -0.1) -- (1.2, 0.1);
\draw[thick] (0, -0.1) -- (0, 0.1);  
\draw[thick] (0, 0.7) node[above, yshift=2pt] {1} -- (1.2, 0.7) node[above, yshift=2pt] {$\infty$};
\draw[thick] (1.2, 0.6) -- (1.2, 0.8);  
\draw[thick] (0, 0.6) -- (0, 0.8);
\draw (0.2,0) arc (-180:-240:0.8);
\draw (1,0) arc (360:420:0.8);
\draw (1,0) arc (-20:-160:0.43);
\node at (0.2, 0) [dcirc] {};
\node at (1, 0) [dcirc] {};
\node at (0.6, 0.7) [dcirc] {};
\end{tikzpicture} &~=~ -8g_0^3\int_1^\infty\int_0^z\int_0^{x_2} \frac{\ud x_1\,\ud x_2\,\ud x_3}{(x_{12}x_{13}x_{23})^{1-\epsilon}} \\
&~=~ \frac{8g_0^3}{\epsilon(1-2\epsilon)(1-3\epsilon)}\int_0^z \ud x_2\, x_2^\epsilon(1-x_2)^\epsilon \left[\begin{tabular}{rl}
&\hspace{-0.4cm} $\epsilon\ {}_2F_1(1-\epsilon,1,2-2\epsilon,x_2)$ \\
  $-$ &\hspace{-0.4cm} $(1-2\epsilon){}_2F_1(1-\epsilon,1,1+\epsilon,x_2)$
\end{tabular}\right] \\
&~=~-8g_0^3\int_0^z \ud x_2\, \left[\frac 1{\epsilon(1-x_2)} +\frac{3x_2+x_2\log x_2+(1+2x_2)\log(1-x_2)}{x_2(1-x_2)}\right] \\
&~=~ 8g_0^3\left[\frac{\log(1-z)}\epsilon +3\log(1-z)+\frac32\log(1-z)^2+\log z\log(1-z)+2\text{Li}_2(z)\right]\:.
\ea

All in all, we get 
\ba
\langle \mu(\infty) \mu(1) \mu(z)\mu(0)\rangle = \frac{\mathcal{N}_\mu^2}{z^{2\Delta_\mu}(1-z)^{2\Delta_\mu}}\left[1+\frac{\epsilon^3}8\big(\log z \log(1-z)+2\text{Li}_2(z)\big)\right].
\ea

Expanding it in blocks in the two different channels:
\ba
\mu \times \mu &= \mathbf{1}+ \left(\frac{\epsilon^2}8+\frac{\epsilon^3}4\right)\phi^2\\
&= \left(1+\frac{\epsilon^3\pi^2}{24}\right)\mathbf{1}+ \left(\frac{\epsilon^2}8-\frac{\epsilon^3}4\right)\tilde{\phi}^2 \:,
\ea
we obtain a check on $\lambda_{\mu\mu\phi^2}^2$ and $\lambda_{\mu\mu\tilde{\phi}^2}^2$. The normalization of the identity is not the same in the two channels. Their ratio is the $g$-ratio (it matches the non-perturbative result)
\ba
\frac{g_{\tt N}}{g_{\tt D}} = 1 + \frac{\epsilon^3\pi^2}{24} \:.
\ea

\section{Interface energy $\Delta^{\tt N/LRI}_\mu$}
\label{app:dnlri}
We compute the partition function in the presence of the $\phi^4$ insertion $\frac{g_0}{4!} \int_0^T :\phi^4:$\,. 
\ba
\log Z(T)=-\frac{g_0^2 T^{2\epsilon}}{48 \epsilon(1-2\epsilon)}+\frac{g_0^3 T^{3\epsilon}}{24\epsilon(1-3\epsilon)}\,\frac{\Gamma(\epsilon)^2}{\Gamma(2\epsilon)}+\frac{g_0^4 T^{4\epsilon}}{4\epsilon(4\epsilon-1)}\, W_4(\epsilon)\:.
\ea
Note that to cubic order in the coupling the computation only depends on conformal symmetry and hence is essentially identical to the one of section \ref{app:perturbative-dmu} up to different combinatorial factors. The last term contains the integrated connected four-point diagrams and will be explained and explicited below.

The beta function is written as
\ba
\beta(g)=-\epsilon g+3 g^2-6 A g^3+B g^4\,, \qquad A=\pi+4\log(2)\:.
\ea
The three loop beta function was computed by~\cite{Benedetti:2020rrq} albeit in a different renormalisation scheme. To determine $B$ we can use the fact that the correction to scaling exponent $\omega:=\beta'(g^*)$ is scheme independent. We find on the one hand with our scheme,
\ba
\omega=\epsilon-\frac{2A}{3} \epsilon^2+\omega_3 \epsilon^3+O(\epsilon^4)\:,
\ea
with $\omega_3=-\frac{8 A^2}9+\frac{2 B}{27}$. Using the results of~\cite{Benedetti:2020rrq} we find the same expression but where now
\ba
\omega_3=\frac 2{27}\big[\alpha_{I_4}+9\pi^2+78\pi \log(2)+156 \log^2(2)-4\pi\big]\::,
\ea
allowing us to determine $B$. Here $\alpha_{I_4}$ is a $d=1$ tetrahedral integral which reference~\cite{Benedetti:2020rrq} cites as evaluating numerically to $330.0089...$. We point out that for $d=1$ this is actually a Selberg integral which can be performed exactly, giving:
\ba
\alpha_{I_4}=6\frac{\Gamma(\tfrac 14)^4}{\pi}\:, \label{eq:exacti4}
\ea
in agreement with the numerical value (this has also been found in~\cite{Zheng:2026eti}).
Knowing the $\beta$ function we can express the bare coupling $g_0$ in terms of the renormalised one, $g_0=\mu^\epsilon g \, Z(g,\epsilon)$ as usual. We can also determine the fixed point coupling $g_*$. In this way we get:
\ba
\Delta_{\mu}^{\tt N/LRI}=-\frac 12 \frac{\partial \log Z(T)}{\partial \log(\mu T)}\bigg|_{g=g_*}=\frac{\epsilon^2}{324}+\frac{A}{324} \epsilon^3+\left(\frac{4A^2-3 \omega_3+8 w_0+4\pi^2}{1296}\right)\,\epsilon^4\:, \label{dimtemp}
\ea
where
\ba
w_0:=\left[W_4(\epsilon)-\frac{9}{8\epsilon^2}-\frac{A}{4\epsilon}\right]\bigg|_{\epsilon=0}\:.
\ea
Thus we have now determined the result in terms of the finite part of the integrated connected four point function of $\phi^4$, which we will now investigate.

\subsubsection*{The quartic diagrams}
At quartic order in the coupling we can write
\ba
\log Z(T)|_{g_0^4}=g_0^4 T^{4\epsilon}\int_{S_4} \prod_{i=1}^4\ud x_i \langle \cO(x_1)\ldots \cO(x_4)\rangle_c\,, \qquad \cO=\frac{:\phi^4:}{4!} \:, \label{eq:4cumulant}
\ea
where $S_4$ is the simplex $0<x_1<x_2<x_3<x_4<1$.
For a connected graph let us set $m_{ij}$ denote the number of propagators between vertices $i,j$. Taking into account the $1/4!$ factors, the total set of graphs with a given set of $m_{ij}$ contribute a term
\begin{equation}
\frac{1}{\prod_{i<j}m_{ij}!}
\prod_{i<j}|x_i-x_j|^{-2m_{ij}\Delta }\:,\qquad \Delta=\frac{1-\eps}{4}\:.
\end{equation}
Normal ordering imposes $m_{ii}=0$; quartic vertices impose $\sum_{j\ne i}m_{ij}=4$. 
The vertex constraints force opposite-edge multiplicities to be equal:
\begin{equation}
 m_{12}=m_{34}=a,\quad m_{13}=m_{24}=b,\quad m_{14}=m_{23}=c,
 \qquad a+b+c=4\:.
 \label{eq:opposite}
\end{equation}
Discarding disconnected graphs leaves the twelve labeled triples given by the permutations of $(3,1,0)$, $(2,2,0)$ and $(2,1,1)$. 

Let us now set $(x_1,x_2,x_3,x_4)=(t,t+rx,t+ry,t+r)$, with $0<t<1-r$, $0<r<1$ and $0<x<y<1$. Integrating over translation $t$ and scale $r$ yields
\begin{equation}
\int_0^1\ud r\,(1-r)r^{-2+4\eps}
=\frac{1}{4\eps(4\eps-1)}\:.
\end{equation}
The remaining ordered integral is
\begin{equation}
J_{a,b,c}(\eps)=\int_{0<x<y<1}\ud x\ud y\,
\frac{1}{[x(1-y)]^{2a \Delta}[y(1-x)]^{2 b \Delta}(y-x)^{2 c\Delta}}\:.
\label{eq:Jabd}
\end{equation}
Thus
\begin{equation}
\log Z(T)|_{g_0^4}=
\frac{g_0^4 T^{4\eps}}{4\eps(4\eps-1)}W_4(\eps),\qquad
W_4(\eps)=\sum_{\substack{a+b+c=4\\\mathrm{connected}}}
\frac{J_{a,b,c}(\eps)}{(a!b!c!)^2}\:.
\label{eq:Wdef}
\end{equation}

\subsubsection*{Analytic evaluation of finite part}
Let us now examine the integrals $J_{a,b,c}$. 
A useful change of coordinates is
\begin{equation}
 x=\frac{u}{1+u},\qquad y=\frac{v}{1+v},\qquad u=zv,\qquad (v>0,\ 0<z<1)\:.
\end{equation}
Using $a+b+c=4$, the Jacobian and all $(1+u)$, $(1+v)$ factors simplify and we can do one of the integrals to obtain
\begin{equation}
J_{a,b,c}(\eps)=\frac{\Gamma(2\epsilon)^2}{\Gamma(4\epsilon)}\,\int_0^1\ud z\,
 z^{-2a \Delta}(1-z)^{-2c \Delta}
 {}_2F_1(2\epsilon,2\epsilon;4\epsilon;1-z)\:.
\label{eq:Jhyper}
\end{equation}
Now let us note that 
\begin{equation}
\frac{\Gamma(2\epsilon)^2}{\Gamma(4\epsilon)}{}_2F_1(2\epsilon,2\epsilon;4\epsilon;1-z)
=\sum_{n=0}^{\infty}\frac{(2\epsilon)_n^2}{(n!)^2}z^n
\left[2\psi(n+1)-2\psi(n+2\epsilon)-\log z\right].
\label{eq:zero-balanced}
\end{equation}
It is easily seen that for the possible values of $a,b,c$ the $n\geq 1$ terms can be integrated and give irrelevant $O(\epsilon)$ contributions to $J_{a,b,c}$. The $n=0$ term is then integrated and we~find
\begin{equation}
W_4(\eps)=\frac{9}{8\eps^2}
+\frac{\pi+4\log2}{4\eps}+w_0+O(\eps)\:,
\label{eq:Wlaurent}
\end{equation}
with
\begin{equation}
w_0=-\frac{13\pi^2}{24}-\frac{\pi+1}{9}
+\frac12\log^2 2 \:.
\label{eq:w0}
\end{equation}
Plugging this result into \reef{dimtemp} we finally find
\begin{equation}
\boxed{
\begin{aligned}
\Delta_\mu^{\tt N/LRI}={}&\frac{\eps^2}{432}+\frac{\pi+4\log2}{324}\eps^3\\
&+\frac{\eps^4}{11664}\left[15\pi^2+132\pi\log2+300\log^22-8
-12\frac{\G(\tfrac14)^4}{\pi}\right]+O(\eps^5)\:.
\end{aligned}
}
\end{equation}

\section{Interface energy $\Delta^{\tt D/LRI}_\mu$}
\label{sec:deltaDlri}
We compute the partition function in the presence of the $\phi^4$ insertion $\frac{g_0}{4!} \int_0^T :\phi^4:$ and a pair of $\tt N/D$ interfaces at $0$ and $T$:
\ba
\log[Z(T) T^{2\Delta_\mu^{\tt N/D}}]=-g_0 \lambda^{\mu \mu}_{\cO} \frac{\Gamma(\epsilon)^2}{\Gamma(2\epsilon)}\, T^{\epsilon}+ \frac{g_0^2}{4!}\, T^{2\epsilon} I_2(\epsilon)+O(g_0^3)\:,
\ea
with
\ba
I_2=\frac{1}{4!}\int_0^T \ud x \int_0^x \ud y\, T^{2\Delta_{\mu}^{\tt N/D}-2\epsilon} \langle \mu(T) :\!\phi^4(x)\!::\!\phi^4(y)\!: \mu(0)\rangle_c\:,
\ea
with the $c$ subscript standing for connected.

Before proceeding with the computation of $I_2$ let us do some bookkeeping. Let us write
\ba
I_2(\epsilon)=\frac{a_2}{\epsilon^2}+\frac{b_2}{\epsilon}+c_2+O(\epsilon)\:,
\ea
and recall once more that the beta function takes the form
\ba
\beta(g)=-\epsilon\,g+3 g^2+\beta_3 g^3+O(g^4)\,, \qquad \beta_3=-6(\pi+4\log(2))\:,
\ea
so that
\ba
g_0=g+\frac{3\,g^2}{\epsilon}\:.
\ea
Let us also write
\ba
\lambda_{\cO}^{\mu \mu}=\frac{\lambda^{\mu \tilde \mu}_{\phi^4}}{\sqrt{24}}=\lambda_0+\epsilon \lambda_1+O(\epsilon^2)\:,
\ea
with
\ba
\lambda_0=\frac{\pi}{\Gamma(\tfrac 14)^4}\,, \quad \lambda_1=\lambda_0(2+\log(2))\:.
\ea
Tuning to the fixed point we find
\ba
 \log[Z(T) T^{2\Delta_\mu^{\tt N/D}}]=-2\gamma_2 \log T+\log N^{\tt D/LRI}\:,
\ea
with $N^{\tt D/LRI}$ a constant depending on $g,\epsilon$ which we will not need explicitly
and
\ba
\gamma_2:=\Delta_\mu^{\tt D/LRI}-\Delta_\mu^{\tt N/D}
=\frac{\lambda_0  \epsilon }{3}+\frac{\epsilon ^2}{216}  \left(-b_2-8\beta_3 \lambda_0+144 \lambda_1\right)+O\left(\epsilon ^3\right)\:.\label{gamma2Dlri}
\ea
More precisely, this expression is valid provided that the following pole cancellation condition is satisfied:
\ba
a_2=72 \lambda_0\:.
\ea

\subsubsection*{Evaluation of $I_2$}
Let us define
\ba
\frac 1{4!}\langle \mu|\phi^4(1) \phi^4(z)|\tilde \mu\rangle_c:=\frac{F(z)}{(1-z)^{2(1-\epsilon)}}\:.
\ea
Note that $F(z)$ can be obtained straightforwardly from \reef{defectcorrs}. Then note that
\ba
\langle \mu(T) :\!\phi^4(x)\!::\!\phi^4(y)\!: \mu(0)\rangle_c= \, \frac{F\left(\tfrac{y(T-x)}{x(T-y))}\right)}{T^{2\Delta_{\mu}^{\tt N/D}}(x-y)^{2(1-\epsilon)}}\:. 
\ea
After some simple changes of variables we can do one of the integrals in $I_2$ to find
\ba
I_2=\frac{\Gamma (2 \epsilon )^2}{\Gamma(4\epsilon)}\int_0^1 \ud z\left(  \, \pFq{2}{1}{2\epsilon,2\epsilon}{4\epsilon}{1-z}\, \frac{F(z)}{(1-z)^{2(1-\epsilon)}}\right).
\ea
The $\epsilon^{-2}$ term can be determined analytically by using that
\ba
\frac{F(z)}{(1-z)^{2(1-\epsilon)}}\supset (1-z)^{\epsilon-1} 72 \lambda_0\:,
\ea
leading to precisely the relation required by the pole cancellation condition. The remaining coefficients in the $\epsilon$ expansion of $I_2$ can only be determined numerically. In particular we~get
\ba
b_2\sim 4.69017\ldots
\ea
Plugging this with the remaining parameters into
\reef{gamma2Dlri} we find the result quoted in the main text:
\ba
\Delta_\mu^{\tt D/LRI} = \frac{1}{64} + \epsilon \left(\frac1{32}+\frac{\pi}{3}\,\frac1{\Gamma\big(\frac14\big)^4}\right) + \epsilon^2\, 0.0504496\ldots +O(\epsilon^3)\:.
\ea

\bibliography{bib}
\bibliographystyle{JHEP}

\end{document}